\documentclass{article}

\usepackage{amssymb}
\usepackage{amsmath}
\usepackage{amsthm}
\usepackage{graphics, graphicx}
\usepackage{color}
\usepackage{epsfig}
\usepackage{enumerate}
\usepackage{caption}
\graphicspath{{./}}

\usepackage[british]{babel}
\usepackage{multicol}
\usepackage{array}
\usepackage{float}
\usepackage{textcomp}
\usepackage{bbm}
\usepackage{mathtools}
\usepackage{framed} 
\usepackage{empheq}
\usepackage{color}
\usepackage[margin=1.4in]{geometry}

\theoremstyle{remark}
\newtheorem{remn}{Remark}

\newcommand{\R}{\mathbb{R}}

\newcommand{\N}{\mathbb{N}}

\newcommand{\eps}{\varepsilon}

\newcommand{\rd}{\mathrm{d}}
\renewcommand{\v}[1]{\mathbf{#1}}

\providecommand{\keywords}[1]{\textbf{\textrm{Keywords---}} #1}
\numberwithin{equation}{section}

\title{Non-local kinetic and macroscopic models for self-organised animal aggregations}

\author{J. A. Carrillo\thanks{Department of Mathematics, Imperial College
London, London SW7 2AZ, UK.}, R. Eftimie\thanks{Division of Mathematics, University of Dundee, Dundee DD1 4HN, UK.}, F. K. O. Hoffmann\thanks{Centre for Mathematical Sciences, University of Cambridge, Cambridge CB3 0WA, UK.}}

\begin{document}

\maketitle

\begin{abstract}
The last two decades have seen a surge in kinetic and macroscopic models derived to investigate the multi-scale aspects of self-organised biological aggregations. Because the individual-level details incorporated into the kinetic models (e.g., individual speeds and turning rates) make them somewhat difficult to investigate, one is interested in transforming these models into simpler macroscopic models, by using various scaling techniques that are imposed by the biological assumptions of the models. Here, we consider three scaling approaches (parabolic, hydrodynamic and grazing collision limits) that can be used to reduce a class of non-local 1D and 2D models for biological aggregations to simpler models existent in the literature. Next, we investigate how some of the spatio-temporal patterns exhibited by the original kinetic models are preserved via these scalings. To this end, we focus on the parabolic scaling for non-local 1D models and apply asymptotic preserving numerical methods, which allow us to 
analyse changes in the patterns as the scaling coefficient $\epsilon$ is varied from $\epsilon=1$ (for 1D transport models) to $\epsilon=0$ (for 1D parabolic models). We show that some patterns (describing stationary aggregations) are preserved in the limit $\epsilon\to 0$, while other patterns (describing moving aggregations) are lost in this limit. To understand the loss of these patterns, we construct bifurcation diagrams. 
\end{abstract}


\section{Introduction}
\label{Sect:introduction}

Over the past 10-20 years a multitude of kinetic and macroscopic models have been introduced to investigate the formation and movement of various biological aggregations: from cells \cite{BellomoDeAngelisPreziosi03_Rev, DeAngelis08_Kinetic_TumorImmun} and bacteria \cite{Pfistner} to flocks of birds, schools of fish and even human aggregations (see, for example, \cite{BellomoVenuti07_PedestrianTraffic_Burger, Carillo_dOrsogna09, DalPasso_Mottoni84_BurgersEq_Population, Carillo2009, DegondMotsch_KineticFish, Betotti08_Kinetic_SocialDynamics, Chertok_PedestrianFlow_Numerics} and the references therein).  
The use of kinetic or macroscopic approaches is generally dictated by the problem under investigation: (i) kinetic (transport) models focus on changes in the density distribution of individuals that have a certain spatial position, speed and movement direction (or are in some activity state \cite{Bellomo09_Kinetic_LivingSystem}); (ii) macroscopic models focus on changes in the averaged total density of individuals \cite{Carillo09_Kinetic_Attraction-Repulsion, Eftimie11_JMB}.  Due to their complex structure, the kinetic models are more difficult to investigate. Although progress has been made in the past years, mainly in regard to the existence and stability of various types of solutions exhibited by these models and the asymptotic methods that allow transitions from kinetic (mesoscopic) to macroscopic models (see, for example, \cite{Carillo_dOrsogna09, Burger:Capasso, DegondMotsch_KineticFish, DegondMotsch_OrientationLimitMacro, Bodnar:Velasquez, BertozziCarrilloLaurent, HaTadmor_ParticleKineticHydrodynamic} 
and the references therein), it is still difficult to study the spatial and spatio-temporal aggregation patterns exhibited by the kinetic models. For example, there are very few studies that investigate the types of spatiotemporal patterns obtained with 2D and 3D kinetic models (see the review in \cite{Eftimie11_JMB}).

\begin{figure}[h!]
\begin{center}
\includegraphics[width=5.2in]{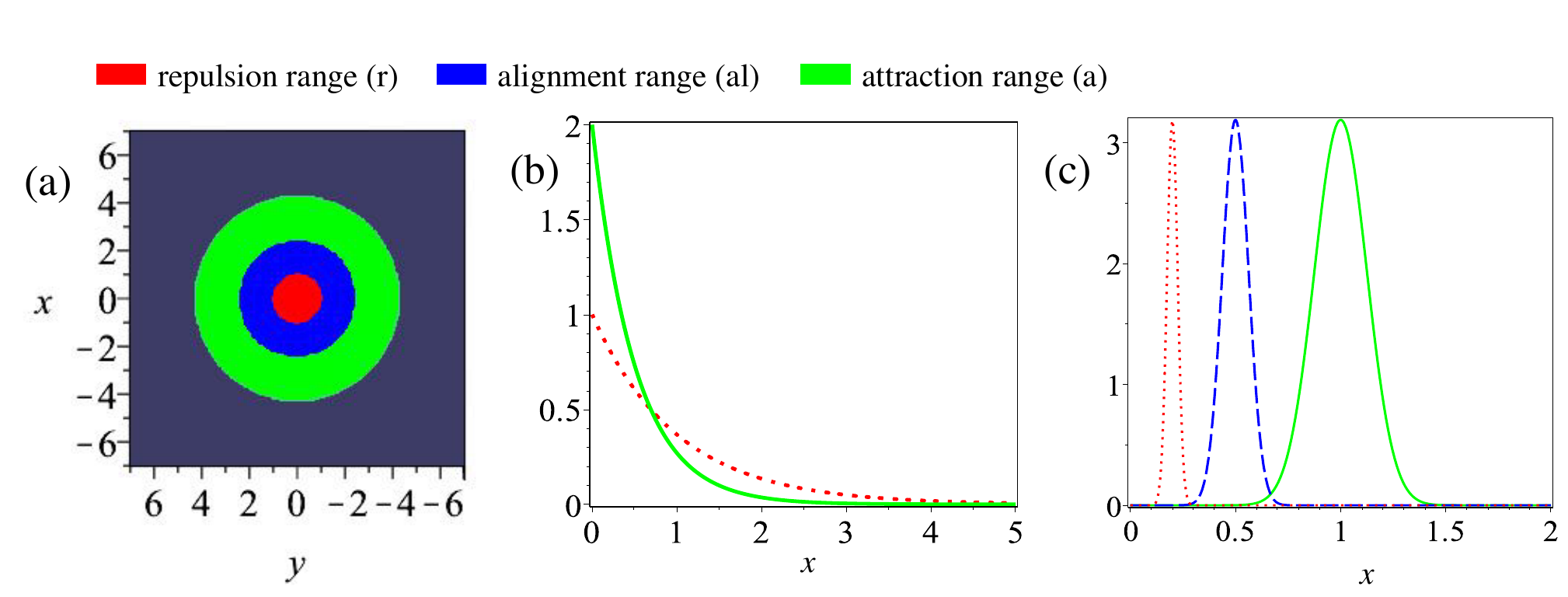}
\caption[]{\footnotesize 2D and 1D spatial kernels for social interactions. (a) 2D: Attractive ($K_{a}$), repulsive ($K_{r}$) and alignment ($K_{al}$) kernels described by equation (\ref{distancekernels}); (b) 1D: Morse-type kernels: $K_{r,a}(x)=e^{-|x|/s_{r,a}}$. (c) 1D: Translated Gaussian kernels: $K_{j}=(1/\sqrt{2\pi m_{j}^{2}})e^{-(x-s_{j})^{2}/(2m_{j}^{2})}$, $m_{j}=s_{j}/8$, $j=r,al,a$. Here, $s_{j}$ denote the mid of the interaction range $j$, while $m_{j}$ controls the width of the interaction range $j$, for $j=r,al,a,$.}
\label{Spatial_ranges}
\end{center}
\end{figure}

Generally, these kinetic and macroscopic models assume that individuals/particles/cells can organise themselves in the absence of a leader. The factors that lead to the formation of these self-organised aggregations are the interactions among individuals as a results of various social forces: repulsion from nearby neighbours, attraction to far-away neighbours (or to roosting areas \cite{CarrilloKlarMartin}) and alignment/orientation with neighbours positioned at intermediate distances. These interaction forces are usually assumed to act on different spatial ranges, depending on the communication mechanisms used by individuals; e.g., via acoustic long-range signals, or via chemical/visual short-range signals. The non-locality of the attractive and alignment/orientation interactions is supported by radar tracking observations of flocks of migratory birds, which can move with the same speed and in the same direction despite the fact that individuals are 200-300 meters apart from each other \cite{Larkin_nonlocal_
birds}. For the repulsive forces some models consider non-local effects generated by decaying interactions with neighbours positioned further and further away \cite{Eftimie2}, while other models consider only local effects \cite{TBL}. In the case of continuous mesoscopic and macroscopic models, these non-local interactions are modelled by interaction kernels (see Figure~\ref{Spatial_ranges} for 2D and 1D kernels). The most common choices for these kernels are Morse potential-type kernels \cite{Carillo09_Kinetic_Attraction-Repulsion, Carillo_dOrsogna09,Carillo2009,CHM2014} (see Figure \ref{Spatial_ranges}(b)) and Gaussian kernels \cite{Eftimie1,Eftimie2,Eftimie11_JMB,M&K} (see Figure \ref{Spatial_ranges}(c)). 

The presence of these non-local interaction terms increases the complexity of the models, rendering them more difficult to be analysed mathematically (see, for example, the 2D model introduced in \cite{Fetecau2010}). It also makes it more difficult to investigate numerically the types of spatial and spatio-temporal patterns exhibited by these models. In general, it is expected that these non-local models would exhibit a variety of patterns. While numerical and analytical studies have been conducted to investigate the patterns in 1D non-local models \cite{Eftimie2, Eftimie3, BE-HH}, such an investigation is still difficult for 2D non-local models, see \cite{Fetecau2010}. 

Another aspect not investigated enough by the existent studies is related to the multi-scale aspect of various self-organised biological aggregations and, in particular, the preservation of patterns between the mesoscopic and macroscopic scales. In the last two decade, various asymptotic methods have been employed to derive macroscopic-level models from kinetic 
models for self-organised biological aggregations (see, for example, \cite{Hillen:Othmer, Othmer:Hillen, Bellomo_hyperbolic07, HaTadmor_ParticleKineticHydrodynamic} and the reference therein). While asymptotic preserving numerical schemes have been derived since late 1990's to investigate the asymptotic dynamics of various transport models \cite{Klar98, Klar99,CarrilloGoudon}, they have only recently been applied to investigate computationally multiscale aspects of biological aggregations \cite{CarrilloYan2013}.

\begin{figure}[h!]
\begin{center}
\includegraphics[width=3.0in]{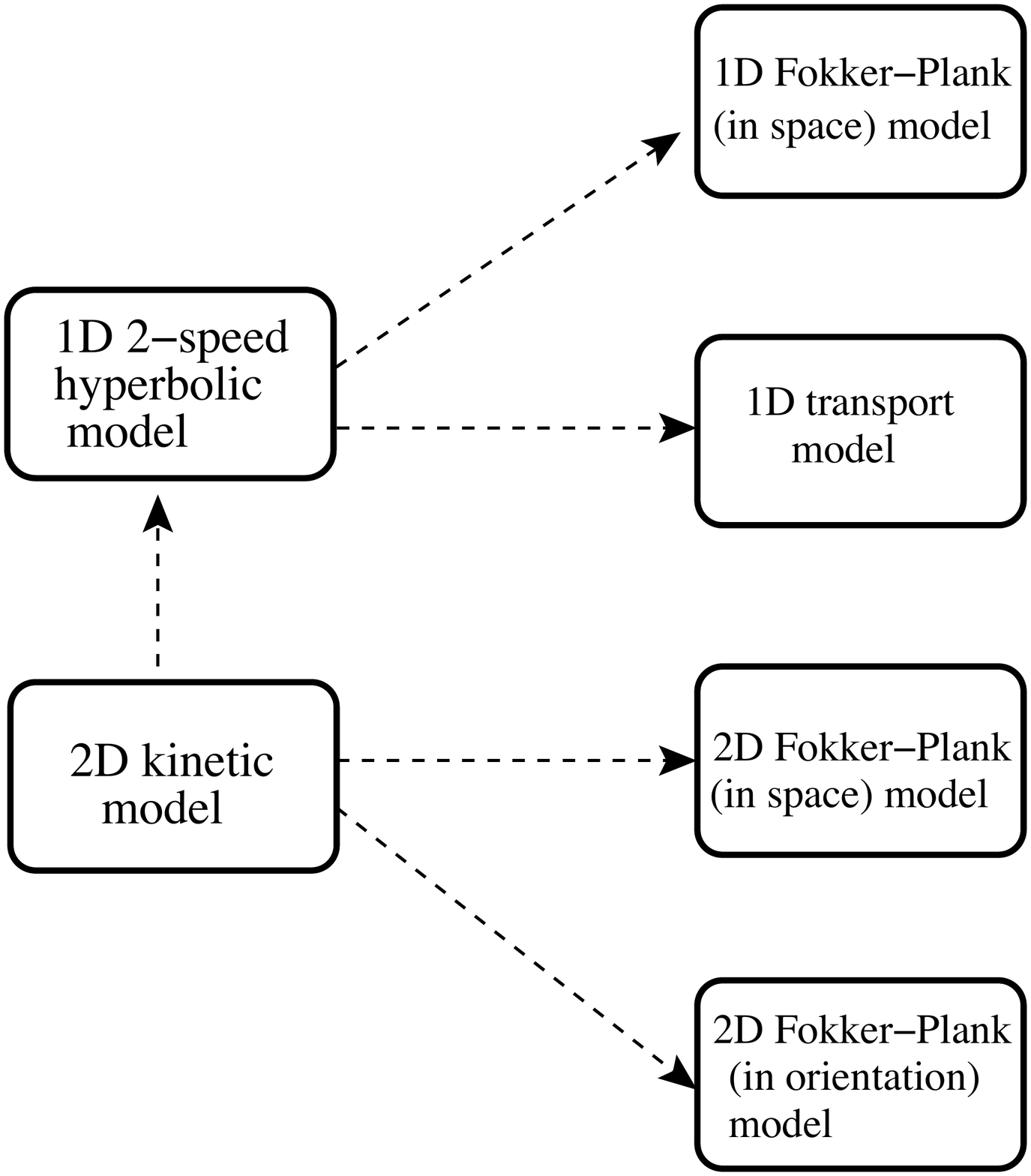}
\caption[]{\footnotesize Schematic diagram of the scaling and reductionist approach taken here.
}
\label{Model_Scheme}
\end{center}
\end{figure}

The goal of this article is to start with a class of 1D and 2D non-local kinetic models for self-organised aggregations that incorporate all three social interactions, and to show, through different scaling approaches, that these models can be reduced to known non-local hyperbolic and parabolic models for swarming; see Figure \ref{Model_Scheme} for a diagram illustrating this approach. Similar scalings have been done in one dimension in the context of bacterial chemotaxis \cite{Saragosti10_Bacteria_TravelPulse} and for the kinetic model (\ref{1D_model}) for individuals moving along a line \cite{Eftimie:thesis}. However, in 2D, the non-local dependence on the three types of social interactions adds another level of complexity compared to the 1D problem. Moreover, to obtain a better understanding of the dynamics of the models following the scaling approach, we also investigate whether various spatial and spatio-temporal patterns exhibited by the  kinetic models are preserved in the limiting parabolic models. 
For simplicity, here we focus 
only on the 1D case. We use asymptotic preserving methods to investigate numerically the preservation of stationary aggregations (that arise via steady-state bifurcations) and moving aggregations (that arise via Hopf bifurcations), as the scaling parameter $\epsilon$ is varied from large positive values ($\epsilon=1$) corresponding to the kinetic models to zero values corresponding to the limiting parabolic models. To show the transitions between different patterns as $\epsilon \to 0$, we construct bifurcation diagrams for the amplitude of the solutions.


The article is structured as follows. Section \ref{1D_hyperbolic} contains a detailed description of the 1D non-local models for animal aggregations, followed by the parabolic and hyperbolic scaling of these models. Section \ref{model_description} contains a description of the 2D non-local models, followed by a parabolic limit and a ``grazing collision" limit, which lead to different types of macroscopic models of parabolic type. Section \ref{Sect:asymptotic} focuses on asymptotic preserving methods for 1D models, and shows the spatial and spatio-temporal patterns obtained with the parabolic and kinetic models, for some specific parameter values. We conclude in Section \ref{Sect:discussion} with a summary and discussion of the results.
  
\section{Description of 1D models}\label{1D_hyperbolic}

The following one-dimensional model was introduced in \cite{Eftimie1,Eftimie2} to describe the movement of the densities of left-moving ($u^{-}$) and right-moving ($u^{+}$) individuals that interact with conspecifics via social interactions:
\begin{subequations}
\label{1D_model}
\begin{align}
\frac{\partial u^{+}}{\partial t}+\gamma \frac{\partial u^{+}}{\partial x}&=-u^{+}\lambda^{+}[u^{+},u^{-}]+u^{-}\lambda^{-}[u^{+},u^{-}],\label{1D_model;a}\\
\frac{\partial u^{-}}{\partial t}-\gamma \frac{\partial u^{-}}{\partial x}&=u^{+}\lambda^{+}[u^{+},u^{-}]-u^{-}\lambda^{-}[u^{+},u^{-}].\label{1D_model;b}
\end{align}
\end{subequations}
Here $\gamma$ is the constant speed and $\lambda^{\pm}$ are the density-dependent turning rates. To model these turning rates, we recall the observation made by Lotka \cite{Lotka}: ``the type of motion presented by living organisms ... can be regarded as containing both a systematically directed and also a random component". We assume that:
\begin{itemize}
\item individuals can turn randomly at a constant rate approximated by $\lambda_{1}$ \cite{Eftimie1}; 
\item individuals can turn randomly in response to the perception of individuals inside any of the repulsive/attractive/alignment ranges (and independent of the movement direction of these neighbours). We approximate this turning rate by $\lambda_{2}$. The non-directed interactions with neighbours are described by the term $y_{N}[u]$. Note that this assumption was not considered previously in \cite{Eftimie1,Eftimie2}, but its necessity will become clearer in Section \ref{model_description}, when we discuss the corresponding 2D model.
\item individuals can turn in response to interactions with neighbours positioned within the repulsive ($r$), attractive ($a$) and alignment ($al$) zones, respectively (see Figure \ref{Spatial_ranges}(a)) \cite{Eftimie1}. This turning is directed towards or away from neighbours, depending on the type of interaction (attractive or repulsive). For alignment interactions, individuals turn to move in the same direction as their neighbours. We can approximate this directed turning rate by $\lambda_{3}$. The non-local directed interactions with neighbours are described by terms $y_{D}^{\pm}[u^{+},u^{-}]$.
\end{itemize}

All these assumptions are incorporated into the following equation for the turning rates, where we denote $u=u^{+}+u^{-}$ the total population density:

\begin{align}
\lambda^{\pm}[u^{+},u^{-}]=&\lambda_{1}+ \lambda_3 \Big( \lambda_{2} f(y_{N}[u]) +  f(y_{D}^{\pm}[u^{+},u^{-}]) \Big).
\label{eq_lambda}  
\end{align}   
The turning function $f(\cdot)$ is a nonnegative, increasing, bounded functional of the interactions with neighbours. An example of such function is $f( Y)=0.5+0.5\tanh(Y-y_{0})$  (see \cite{Eftimie2}), where $y_{0}$ is chosen such that when $Y=0$ (i.e., no neighbours around), then $f(0)\approx 0$ and the turning is mainly random. 
For the non-directed density-dependent turning we define the turning kernel $K^{N} = q_{r}K_{r} + q_{al}K_{al} + q_{a}K_{a}$ obtained by superimposing the kernels for the repulsion  ($K_{r}$), alignment ($K_{al}$) and attraction ($K_{a}$) ranges (see Figure \ref{Spatial_ranges} for examples of such kernels in 1D and 2D). Here $q_{r}$, $q_{al}$ and $q_{a}$ represent the magnitudes of the social interactions on each of these ranges.  Note that in \cite{Eftimie2}, $\lambda_{2}=0$ and this density-dependent random turning term does not exist. However, in 2D, this term appears naturally when we incorporate random turning behaviour (as discussed in Section \ref{model_description}). 

With these notations we may define 
\begin{equation}
y_{N}[u]=K^{N}\ast u \label{eq_N}
\end{equation}
and
\begin{equation}
y_{D}^{\pm}[u^{+},u^{-}]=y^{\pm}_{r}[u^{+},u^{-}]-y^{\pm}_{a}[u^{+},u^{-}]+y^{\pm}_{al}[u^{+},u^{-}], \label{eq_y}
\end{equation}
for the non-directed and directed density-dependent turning mechanisms, respectively. Here, $y_{j}^{\pm}[u^{+},u^{-}]$, $j=r,al,a,$ describe the directed turning in response to neighbours within the repulsive ($r$), alignment ($al$) and attractive ($a$) social ranges (as in \cite{Eftimie1}). In contrast to $K^N \ast u$ (where individuals turn randomly whenever they perceive other neighbours around), here the direction of the turning is given by incorporating movement direction: towards or away conspecifics. 
For this reason,  $y_{a}^{\pm}$ and $y_{r}^{\pm}$ enter equation (\ref{eq_y}) with opposite signs. Note that (\ref{eq_N}) is one of the possible definitions of $y_{N}[u]$;  depending on how we define the interactions between individuals other definitions make sense too, as we discuss next.

\begin{figure}[h!]
\begin{center}
\includegraphics[width=5.0in]{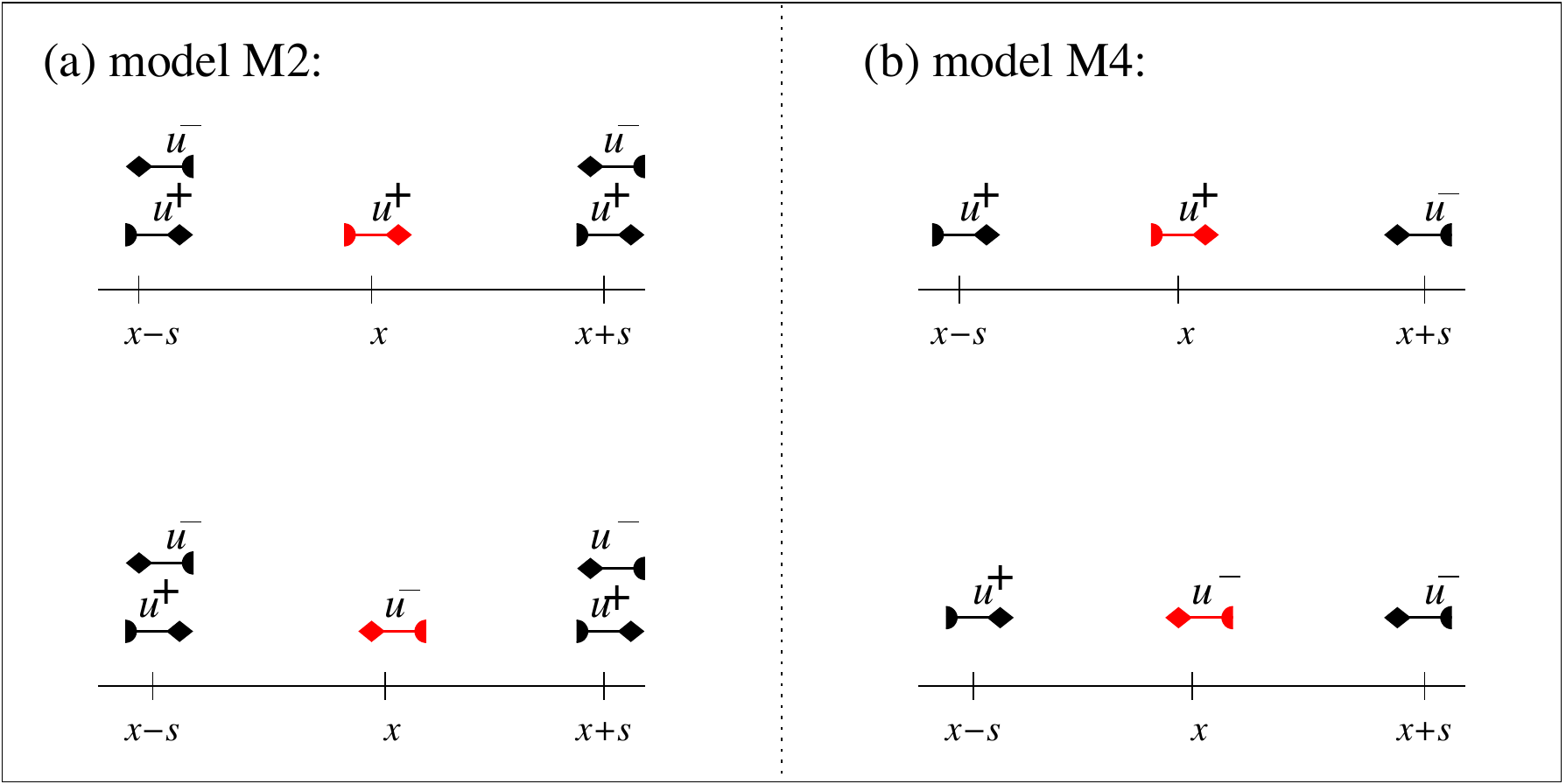}
\caption[]{\footnotesize Diagram describing the mechanisms through which a reference individual positioned at $x$ (right-moving -- top; left-moving -- bottom) perceives its neighbours positioned at $x-s$ and $x+s$. The reference individual can perceive (a) all its neighbours (model M2 in \cite{Eftimie2}); (b) only its neighbours moving towards it (model M4 in \cite{Eftimie2}).
}
\label{Models_M2_M4_Description}
\end{center}
\end{figure}

The density-dependent turning (both non-directed and directed) depends also on how individuals communicate with each other, namely whether they can emit/perceive signals to/from \emph{all} or \emph{some} of their neighbours. Here, we consider two particular situations described by models called M2 and M4, introduced in \cite{Eftimie2}; see Figure \ref{Models_M2_M4_Description}:

\begin{itemize}
\item Individuals communicate via omni-directional communication signals, and thus they can perceive \emph{all} their neighbours positioned around them within all social interaction ranges. For instance, the majority of mammals communicate via a combination of visual, chemical and auditory signals, which allows them to receive/send information from/to all their neighbours.  With this assumption (which corresponds to model M2 in \cite{Eftimie2}; see also Figure~\ref{Models_M2_M4_Description}(a)), the terms $y_{r,a,al}^{\pm}$ are defined as follows:

\begin{subequations}
\label{1D_model_M2y} 
\begin{align}
y_{r,a}^{\pm}= q_{r,a}\int_{-\infty}^{\infty}K_{r,a}(s)\big(u(x\pm s)-u(x\mp s) \big)&ds, \\
y_{al}^{\pm}=q_{al}\int_{-\infty}^{\infty}K_{al}(s)\big(u^{\mp}(x\mp s)+u^{\mp}(x\pm s)&\\
 -u^{\pm}(x\mp s)-u^{\pm}(x\pm s) \big)&ds.\notag
\end{align}
\end{subequations}
Here, $q_{r}$, $q_{al}$ and $q_{a}$ describe the magnitude of the repulsive, alignment and attractive interactions, respectively. $K_{r}$, $K_{al}$ and $K_{a}$ are the social interaction kernels mentioned before and described in Figure \ref{Spatial_ranges}. For comparison purposes, throughout this article we will consider translated Gaussian kernels as in Figure \ref{Spatial_ranges}(c). 

Note that in equations (\ref{1D_model_M2y}) we have $y^{-}_{j}=-y^{+}_{j}$, $j=r,al,a$. Moreover, for this model, the directionality of neighbours influences only the alignment interactions (the attractive and repulsive interactions being defined in terms of the total density $u=u^{+}+u^{-}$). Also, for this particular model, the random density-dependent terms are given by
\begin{equation}\label{1D_model_M2yN}
y_{N}[u]=\int_{-\infty}^{\infty}K^N(s)\big(u(x+s)+u(x-s))ds.
\end{equation}

\item Individuals communicate via unidirectional communication signals, and thus they can perceive only those neighbours moving towards them. For example, birds communicate via directional sound signals, and to ensure an effective transmission of their signals they orient themselves towards their targeted receivers \cite{BreitwischWhitesides_DirectionalitySound}. With this assumption (which corresponds to model M4 in \cite{Eftimie2}; see also Figure~\ref{Models_M2_M4_Description}(b)), the terms $y_{r,a,al}^{\pm}$ are defined as follows:
\begin{equation}
y_{r,a,al}^{\pm}= q_{r,a,al}\int_{-\infty}^{\infty}K_{r,a,al}(s)\big(u^{\mp}(x\pm s)-u^{\pm}(x\mp s) \big)ds.\label{1D_model_M4y} 
\end{equation}
As before $y^{-}_{j}=-y^{+}_{j}$, $j=r,al,a$. Note that for this model, the directionality of neighbours influences all three social interactions. Moreover, for this model, the random density-dependent terms are given by
\begin{equation}\label{1D_model_M4yN}
y_{N}[u]=\int_{-\infty}^{\infty}K^N(s)\big(u^{-}(x+s)+u^{+}(x-s))ds.
\end{equation}

\end{itemize}
 
We decided to focus on these two particular models because: (i) model (\ref{1D_model})+(\ref{1D_model_M2y})+(\ref{1D_model_M2yN}) assuming $\lambda_1=0$ has been generalised to 2D; (ii) model (\ref{1D_model})+(\ref{1D_model_M4y})+(\ref{1D_model_M4yN}) assuming $\lambda_2=0$ has been investigated analytically and numerically, and showed that it can exhibit Hopf bifurcations (even when $q_{al}=0$), which give rise to spatio-temporal patterns such as rotating waves and modulated rotating waves \cite{BE-HH}. In contrast, model (\ref{1D_model})+(\ref{1D_model_M2y})+(\ref{1D_model_M2yN}) with $\lambda_2=0$ does not seem to exhibit rotating waves when $q_{al}=0$, see \cite{Eftimie2}. Throughout this article we will focus on the $q_{al}=0$ case since, as we will show in the next section, the parabolic scaling in 1D leads to the elimination of $q_{al}$ terms. We will return to these spatio-temporal patterns in Section \ref{Sect:Sim}, when we will investigate numerically the preservation of these patterns in the 
parabolic scaling.   

To complete the description of the model, we need to specify the domain size and the boundary conditions. Throughout most of this article, we will consider an infinite domain. However, for the purpose of numerical simulations, in Sections \ref{Sect:SS_stability} and \ref{Sect:asymptotic} we will consider a finite domain of length $L$  (i.e., $[0,L]$) with periodic boundary conditions:
\begin{equation}
u^{+}(L,t)=u^{+}(0,t),\;\;\; u^{-}(0,t)=u^{-}(L,t).
\end{equation}
This assumption will also require wrap-around conditions for the kernels describing the nonlocal social interactions. (We will return to this aspect in Section \ref{Sect:asymptotic}). For large $L$, this assumption approximates the dynamics on an infinite domain.

In the following, we show how this hyperbolic 2-velocity model can be reduced to other non-local hyperbolic and parabolic models for the density of self-organised aggregations, by considering suitable scalings. The scaling of the variables and parameters of the models depends on the biological phenomena described, as well as on the detailed biological assumptions incorporated into the model. Of course, to be useful in practice, these parameters have to be callibrated and adapted to particular species as in \cite{Hemelrijk:Kunz2004,HCH}. These scaling arguments are classically obtained by writing a dimensionless formulation of the problem. We refer to \cite{Saragosti10_Bacteria_TravelPulse} in bacterial chemotaxis and \cite{ACGS2001} in semiconductor modelling for these detailed standard scaling computations. After this dimensionless rescaling, we typically end up with two different time scales whose balance determines our small parameter: the drift time and the diffusion time.

We start in Subsections \ref{Parabolic_limit} and \ref{Parabolic_limit_linear} with a parabolic scaling, which describes the situation where the drift time of a population is much smaller than its diffusion time, as in \emph{E. coli} bacteria \cite{Hillen:Othmer}. To this end, we discuss two separate cases, which lead to two different parabolic equations. In Subsection \ref{Parabolic_limit} we focus on the case where the social interactions are described by the non-linear function $f(y)$ in \eqref{eq_lambda}. In Subsection \ref{Parabolic_limit_linear}, we focus on the case where the social interactions are described by a linear function $f(y)=y$ in \eqref{eq_lambda}. Finally, in Subsection \ref{Sect:hydrodynamic} we consider a hydrodynamic scaling, which describes the situation where the drift time and the diffusion time have similar magnitudes, as in some cell movement models \cite{Perthame}. 

\subsection{Parabolic limit for non-linear interactions} \label{Parabolic_limit}

To transform the hyperbolic system (\ref{1D_model}) into a parabolic equation, a scaling argument is applied \cite{Hillen:Othmer}. One can scale the space and time variables ($x=x^{*}/\eps$, $t=t^{*}/\eps^{2}$, with $\eps \ll1$), or can scale the speed ($\gamma$) and the turning rates ($\lambda_{1,2,3}$). In both cases, we consider the rescaled interaction kernels $K^\eps_j (x) = \eps K_j(\eps x)$ in the expressions for $y^{\pm}_j$, $j = r, al, a$. To be consistent with the approach in Section \ref{Sect:ParabLimit2D}, here we scale the time and space variables. As mentioned above, the scaling parameter $\epsilon$ depends on the biological problem modelled. For example, in \cite{Hillen:Othmer} the authors connect $\epsilon$ to the ratio of the drift ($\tau_{drift}$) and diffusion ($\tau_{diff}$) times observed in bacteria such \textit{E. coli}, where $\tau_{drift}\approx 100$ seconds and $\tau_{diff} \approx 10^{4}$ seconds, and thus $\epsilon \approx O(10^{-2})$. Similar scaling arguments are used in \cite[
Appendix]{Saragosti10_Bacteria_TravelPulse} to analyse the ability of parabolic scalings to describe travelling pulses.
 
First, let us re-write model (\ref{1D_model}) in terms of the total density $u(x,t)$ and the flux $v(x,t)=\gamma(u^{+}(x,t)-u^{-}(x,t))$ of individuals (see also \cite{Hillen:Othmer,Holmes}):
\begin{subequations}
\label{Cattaneo_Syst}
\begin{align}
\epsilon^{2} \frac{\partial u}{\partial t}+\epsilon \frac{\partial v}{\partial x}&=0,\\
\epsilon^{2} \frac{\partial v}{\partial t}+\epsilon \gamma^{2} \frac{\partial u}{\partial x}&=\gamma u\big( \lambda^{-}[u,v]-\lambda^{+}[u,v]\big)- v\big(\lambda^{+}[u,v]+\lambda^{-}[u,v]\big),
\end{align}
\end{subequations}
with initial conditions $u(x,0)=u_{0}(x)$, $v(x,0)=v_{0}(x)$. For clarity, here we dropped the ``$*$" from the rescaled space ($x^{*}$) and time ($t^{*}$) variables. In addition to this scaling, we also assume that  when $\eps\to 0$  it leads to a reduced perception of the surrounding neighbours \cite{Eftimie:thesis}:
 \begin{equation}
 f_{\epsilon}\Big(y_{D}^{\pm}[u,v]\Big)=\eps f\Big(y_{D}^{\pm}\big[u, \int_{\frac{x^{*}}{\eps}}\eps \frac{\partial u}{\partial t^{*}}\big]\Big), \;\; f_{\epsilon}\Big(y_{N}[u]\Big)=\eps f\Big(y_{N}[u]\Big), 
 \label{fassumption}
 \end{equation}
where $f$ enters the turning functions $\lambda^{\pm}$ (\ref{eq_lambda}):
\begin{eqnarray}
\lambda^{+}[\cdot]+\lambda^{-}[\cdot]&=&2\lambda_{1}+2\lambda_{2}\, \lambda_3 \,\epsilon f(y_{N}[\cdot]) + \eps\, \lambda_3 \, \Big(f(y_{D}^{+}[\cdot])+f(y_{D}^{-}[\cdot])\Big),\nonumber\\
\lambda^{-}[\cdot]-\lambda^{+}[\cdot]&=&\lambda_{3}\eps\Big(f(y_{D}^{-}[\cdot])-f(y_{D}^{+}[\cdot])\Big).\nonumber
\end{eqnarray}
By eliminating $v=\eps \int_{x}\frac{\partial u}{\partial t}$ from equations (\ref{Cattaneo_Syst}), 
and taking the limit $\eps \to 0$, we obtain the following parabolic equation
\begin{align}
\frac{\partial u}{\partial t}
= &\frac{\gamma^2}{2\lambda_1}\, \frac{\partial}{\partial x}\left( \frac{\partial u}{\partial x}\right)
-\frac{\lambda_3 \gamma}{2\lambda_1} \frac{\partial}{\partial x} \left( \big( f(y_{D}^{-}[u])-f(y_{D}^{+}[u]) \big) u \right).\label{Eq_parabolic}
\end{align}
We note here that the non-local terms $f(y_{D}^{\pm}[u])$ now depend only on the repulsive and attractive interactions. The reason for this is that the alignment interactions are defined in terms of $u^{\pm}=(u\pm \frac{1}{\gamma} v)/2=0.5(u\pm \frac{1}{\gamma} \int_{x/\eps}\eps^{2} \partial u/\partial t)$. As $\eps\to 0$, the $u$ terms in (\ref{1D_model_M2y}) cancel out, and the integrals approach zero.  Equation (\ref{Eq_parabolic}) can be re-written as 
 \begin{equation}
 \frac{\partial u}{\partial t}= \frac{\partial}{\partial x}\left(D_{0}\, \frac{\partial u}{\partial x}\right)-\frac{\partial}{\partial x}\Big(B_0\, u \, V(u)\Big), \label{Eq_limit_parab}
 \end{equation}
 with diffusion rate $D_{0}= \gamma^2/(2\lambda_{1})$ and drift rate $B_0= \lambda_{3}\gamma/(2\lambda_{1})$. The velocity $V(u)$ depends on the communication mechanism incorporated. For example, for model M2 we have $y^{\pm}_{D}[u]=\pm K \ast u$, and so the velocity is given by
 \begin{equation*}
  V[u]=f\big(- K \ast u \big)-f\big(K\ast u\big) 
\end{equation*}
 where
 \begin{gather}
  \bar K=q_{r}K_{r}-q_{a}K_{a}, \quad 
  \bar K^{\pm} \ast u = \int_{-\infty}^{\infty} \bar K(s)u(x\pm s)ds,\nonumber\\
   K\ast u = \bar{K}^{+}\ast u - \bar{K}^{-}\ast u. \label{Kernels_pm}
 \end{gather}
 For model M4, we have $y^{\pm}_{D}[u]=\pm 0.5 K \ast u$, and so the velocity is quite similar: $V[u]=f\big(- 0.5 K \ast u \big)-f\big(0.5K \ast u\big)$, the factor $0.5$ appearing from $u^{\pm}=0.5(u\pm \frac{1}{\gamma} v)$.  
 
\begin{remn}
We observe that the random density-dependent turning $f(y_{N}[u])$ does not appear in this parabolic limit. This is the result of the scaling assumptions (\ref{fassumption}).
 \end{remn}
 
 \subsection{Parabolic limit with linear interactions}\label{Parabolic_limit_linear}
In the previous subsection, the turning functions $f(\cdot)$ were chosen to be bounded, since individuals cannot turn infinitely fast when subject to very strong interactions with neighbours \cite{Eftimie2,Eftimie3}. However, for simplicity, many models consider linear functions: $f(z)=z$ (see, for example, \cite{M&K, Mogilner:Keshet1996II,Fetecau2010}). Because the 2D kinetic model that we will investigate in Section \ref{model_description} assumes $f$ to be a linear function, with a very weak directed turning behaviour ($\epsilon \lambda_{3}$), we now consider the case $f(y_{N}[u])=y_{N}[u]$ and $f(y^{\pm}_{D}[u])=\epsilon y^{\pm}_{D}[u]$. 

To preserve the total density, we eliminate $v=\int_{x}\epsilon \frac{\partial u}{\partial t}$ from equations (\ref{Cattaneo_Syst}) and, by taking the limit $\epsilon\to 0$, we obtain the following equation:
\begin{equation}
\gamma^{2}\frac{\partial ^{2}u}{\partial x^{2}}=\gamma \lambda_{3} \frac{\partial}{\partial x}\Big(u(y_{D}^{-}[u]-y_{D}^{+}[u]) \Big)+\frac{\partial}{\partial x}\Big(\int_{x}\frac{\partial u}{\partial t}(2\lambda_{1}+2\lambda_{2}\lambda_{3}y_{N}[u]) \Big)\nonumber
\end{equation}
Integrating first with respect to $x$, and then differentiating with respect to $x$ (to solve for $\partial u/\partial x$) we obtain the following parabolic equation with density-dependent coefficients: 
\begin{subequations}
\label{Eq_parabolic_lambda2}
\begin{align}
& \frac{\partial u}{\partial t}
= \frac{\partial}{\partial x}\left( D[u] \frac{\partial u}{\partial x}\right)-\frac{\partial}{\partial x} \Big(B[u] u\big(y_{D}^{-}[u]-y_{D}^{+}[u]\big) \Big), \\
&D[u]=\frac{\gamma^2}{2(\lambda_1+\lambda_{2}\lambda_{3}K^{N}\ast u)} \;\; \text{and}\;\; B[u]=  \frac{\lambda_3 \gamma}{(2\lambda_1+2\lambda_{2}\lambda_{3} K^{N}\ast u)}.  
\end{align}
\end{subequations}
This expression is similar to the asymptotic parabolic equation (\ref{macromodel}) for the 2D model. We will return to this aspect in Section \ref{Sect:ParabLimit2D}.
 
\subsection{The preservation of steady states and their stability as $\epsilon \to 0$}\label{Sect:SS_stability}

The spatially homogeneous steady states describe the situation where individuals are evenly spread over the whole domain. In the following we investigate how these steady states and their linear stability are preserved in the parabolic limit. To this end, we focus on the more general case of non-linear social interactions (the case with linear interactions is similar). For simplicity we assume here that $\lambda_{2}=0$ and $q_{al}=0$. 

\begin{figure}[!h]
\centering
\includegraphics[width=5.4in]{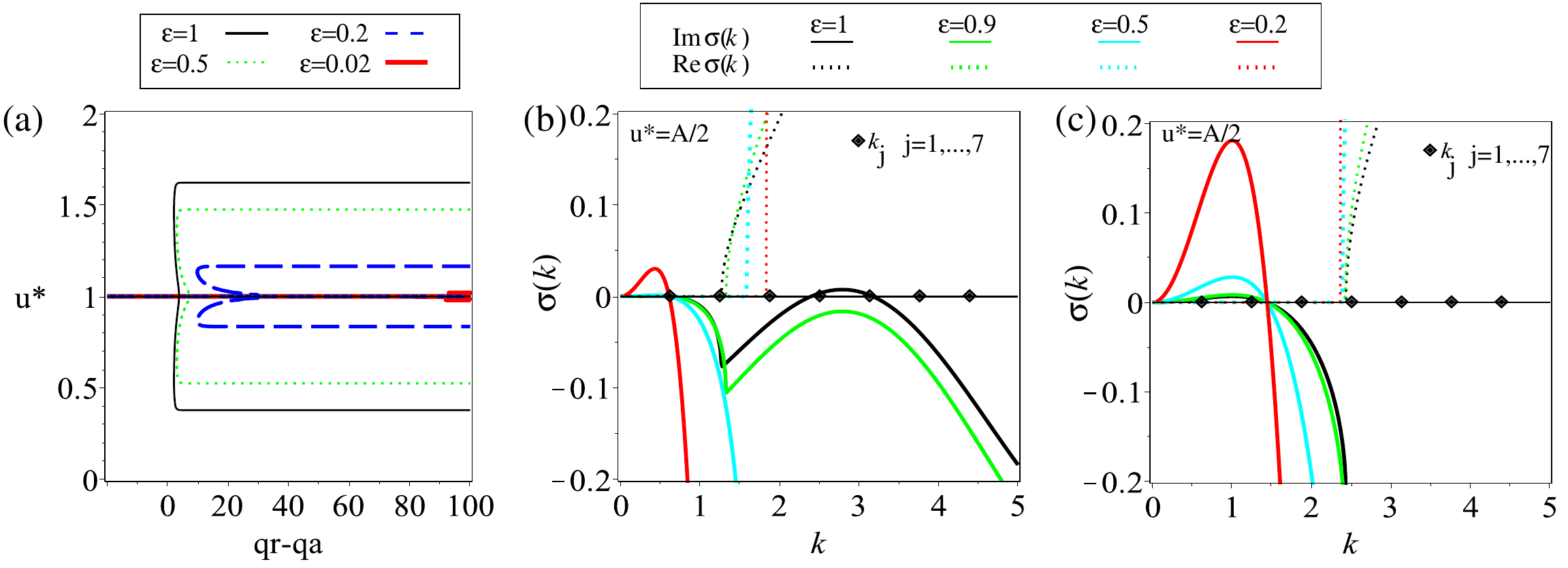}
\caption{(a) Spatially homogeneous steady states $u^{*}$ for model (\ref{Cattaneo_Syst}) with communication signals (\ref{1D_model_M4y}) and (\ref{1D_model_M4yN}) (communication mechanism M4), for different values of $\epsilon$; (b) Dispersion relation $\sigma(k_{j})$ for M4 (given by (\ref{dispersionM4})), showing the stability of the spatially homogeneous steady state $u^{*}=A/2$, for different values of $\epsilon$; (c) Dispersion relation $\sigma(k_{j})$ for M2, for the stability of the spatially homogeneous steady state $u^{*}=A/2$, for different values of $\epsilon$. The continuous curves describe $\text{Re}\: \sigma(k_{j})$, while the dotted curves describe the $\text{Im}\: \sigma(k_{j})$. The small diamond-like points show the discrete wavenumbers $k_{j}, j=1,...,7$. The parameter values are: (b)$q_{a}=1.545$, $q_{r}=2.779$; (c) $q_{a}=1.5$, $q_{r}=0.93$. The rest of parameters are: $q_{al}=0$, $\lambda_{1}=0.2$, $\lambda_{2}=0$, $\lambda_{3}=0.9$, $A=2$.}
\label{Fig_Ss_M4_Epsilon}
\end{figure}

 Figure \ref{Fig_Ss_M4_Epsilon}(a) shows the number and magnitude of the steady states $u^{*}$ displayed by (\ref{Cattaneo_Syst})-(\ref{fassumption}) with communication mechanisms M4, as $\epsilon \to 0$. For  $\epsilon=1$, the model can display up to 5 different steady states: one ``unpolarised" state $(u^{+},u^{-})=(u^{*},u^{*})=(A/2,A/2)$ (where half of the individuals are facing left and half are facing right), and four ``polarised" states $(u^{*},A-u^{*}), (A-u^{*},u^{*})$ characterised by $u^{*}<A/2$ or $u^{*}>A/2$. Here $A$ is the total population density. As $\epsilon$ decreases, the magnitude of the polarised states decreases (i.e., the differences between the number of individuals facing right and those 
facing left are decreasing). Moreover, for small $\epsilon$, these polarised states appear for larger values of $q_{r}-q_{a}$. When $\epsilon=0$ there is only one steady state $u^{*}=A/2$. Since this state exists for all $\epsilon \geq 0$, from now on we will focus our attention only on it. 
Note that, for the communication mechanism M2 (not shown here), when $q_{al}=0$ the nonlocal attractive-repulsive terms vanish, and there is only one steady state, $u^{*}=A/2=1$, which does not depend on $\epsilon$.

Models (\ref{1D_model}) and (\ref{Cattaneo_Syst}) could exhibit a large variety of local bifurcations: codimension-1 Steady-state and Hopf bifurcations \cite{Eftimie3} as well as codimension-2 Hopf/Hopf, Hopf/Steady-state and Steady-state/Steady-state bifurcations \cite{BE-HH}. Next we focus on the parameter region where two such bifurcations can occur. We choose, for example, a Hopf/Steady-state bifurcation for M4 (Figure \ref{Fig_Ss_M4_Epsilon}(b)) and a steady-state bifurcation for M2 (Figure \ref{Fig_Ss_M4_Epsilon}(c)), and investigate what happens with these particular bifurcations when $\epsilon \to 0$. To identify the parameter regions where these bifurcations occur, consider a finite domain of length L, and investigate the growth of small perturbations of spatially homogeneous solutions, i.e., assume $u^{\pm}\propto u^{*}+a_{\pm}\text{exp}(\sigma t+ik_{j}x)$, with $k_{j}=2\pi j/L, j\in\mathbb{N^{+}}$, the discrete wave-numbers, and $|a_{\pm}|\ll1$. We substitute these solutions into the linearised 
system (\ref{Cattaneo_Syst}), and solve for $\sigma$ -- which describes the growth/decay of the perturbations -- as a function of the wave-numbers $k_{j}$.

Figure \ref{Fig_Ss_M4_Epsilon}(b) shows the stability of the spatially homogeneous steady state $u^{*}=A/2$, for model M4, as given by the dispersion relation $\sigma(k_{j})$: 
\begin{equation}
\epsilon^{2}\sigma^{2}+\sigma(2L_{1}^{\epsilon}-R_{2}^{\epsilon}\text{Re}(\hat{K}^{+}))+\gamma^{2}k_{j}^{2}-\gamma k_{j} R_{2}\text{Im}(\hat{K}^{+})=0, \label{dispersionM4} 
\end{equation}
with $L_{1}^{\epsilon}=\lambda_{1}+\epsilon \lambda_{2}f(0)$, $R_{2}^{\epsilon}=2\epsilon u^{*}\lambda_{3}f'(0)$, and $\hat{K}^{+}=\text{Re}(\hat{K}^{+})+i\text{Im}(\hat{K}^{+})$ the Fourier transforms of $\bar{K}^{+}\ast u$ described in \eqref{Kernels_pm}. 
As shown in Figure \ref{Fig_Ss_M4_Epsilon}(b), for $q_{a}=1.545$, $q_{r}=2.779$, $\lambda_{1}=0.2$, $\lambda_2=0$, $\lambda_{3}=0.9$ and $\epsilon=1$, three modes become unstable  at the same time: a steady-state mode $k_{1}$ (associated with stationary patterns) and two Hopf modes $k_{4}$ and $k_{5}$ (associated with travelling patterns). As $\epsilon \to 0$, the steady-state mode persists while the Hopf modes disappear (i.e., Re$(\sigma (k_{4,5}))<0$  and the Hopf modes become stable; see Figure \ref{Fig_Ss_M4_Epsilon}(b).) This can be observed also from equation (\ref{dispersionM4}): as $\epsilon \to 0$, we have $\sigma\in \mathbb{R}$. A similar investigation of the local stability of the spatially homogeneous steady states associated with the non-local parabolic equation (\ref{Eq_limit_parab}) shows that this equation cannot have complex eigenvalues, and thus cannot exhibit local Hopf bifurcations \cite{BE:ParabLimit}.

Figure \ref{Fig_Ss_M4_Epsilon}(c) shows the stability of the spatially homogeneous steady state $u^{*}=A/2$, for model M2, as given by the dispersion relation $\sigma(k_{j})$:
  \begin{equation*}
 \epsilon^{2}\sigma^{2}+\sigma (2L_{1}^{\epsilon})+\gamma^{2}k_{j}^{2}-2\gamma k_{j} R_{2} \text{Im}(\hat{K}^{+})=0. \label{dispersionM2}
 \end{equation*}
 For $q_{a}=1.5$, $q_{r}=0.93$, $\lambda_{1}=0.2$, $\lambda_2=0$, $\lambda_{3}=0.9$ and $\epsilon=1$, two steady-state modes are unstable at the same time: $k_{1}$ and $k_{2}$ (both associated with stationary patterns). As $\epsilon\to 0$, these two modes remain unstable. Hence, we expect that the spatial patterns generated by these modes will persist as $\epsilon \to 0$. 
We will return to this aspect in Section \ref{Sect:Sim}, when we will investigate numerically the mechanisms that lead to the disappearance of the Hopf modes and the persistence of the steady-state modes, as $\epsilon \to 0$. 

\subsection{Hydrodynamic scaling}\label{Sect:hydrodynamic}
We focus again on the more general case of non-linear social interactions $f(y)$. Consider now the following scaling for the space and time variables: $x=x^{*}/\eps$ and $t=t^{*}/\eps$, where $\eps\ll1$ is a small parameter that measures the difference between the time scale for the turning behaviour and the time scale for the changes in population density. 

Then, after dropping the ``*" for clarity, system (\ref{1D_model}) can be written as 
\begin{subequations}
\label{Eq_hyp_scaled}
\begin{align}
\frac{\partial u_{\eps}^{+}}{\partial t}+\gamma \frac{\partial u_{\eps}^{+}}{\partial x}&=\frac{1}{\eps}\Big(-\lambda^{+}[u_{\eps}^{+},u_{\eps}^{-}]u_{\eps}^{+}+\lambda^{-}[u_{\eps}^{+},u_{\eps}^{-}]u_{\eps}^{-} \Big),\label{Eq_hyp_scaled;a}\\
\frac{\partial u_{\eps}^{-}}{\partial t}-\gamma \frac{\partial u_{\eps}^{-}}{\partial x}&=\frac{1}{\eps} \Big(\lambda^{+}[u_{\eps}^{+},u_{\eps}^{-}]u_{\eps}^{+}-\lambda^{-}[u_{\eps}^{+},u_{\eps}^{-}]u_{\eps}^{-} \Big).\label{Eq_hyp_scaled;b}
\end{align}
\end{subequations}
Adding and subtracting the equations in (\ref{Eq_hyp_scaled}) leads to a system similar to (\ref{Cattaneo_Syst}):
\begin{align}
\frac{\partial u_{\eps}}{\partial t}+\frac{\partial v_{\eps}}{\partial x}&=0,\label{Cattaneo_epsilon;a}\\
\eps\frac{\partial v_{\eps}}{\partial t}+\eps\gamma^{2} \frac{\partial u_{\eps}}{\partial x}&=\gamma u_{\eps}\big( \lambda^{-}[u_{\eps},v_{\eps}]-\lambda^{+}[u_{\eps},v_{\eps}]\big)-v_{\eps}\big(\lambda^{+}[u_{\eps},v_{\eps}]+\lambda^{-}[u_{\eps},v_{\eps}]\big),\nonumber
\end{align}
where $u_{\eps}=u^{+}_{\eps}+u^{-}_{\eps}$ and $v_{\eps}=\gamma(u^{+}_{\eps}-u^{-}_{\eps})$. Consider now the expansions 
\begin{eqnarray}
u_{\eps}&\approx& u_{0}+\eps u_{1}+\eps^{2}u_{2}+O(\eps^{3}),\nonumber\\
v_{\eps}&\approx& v_{0}+\eps v_{1}+\eps^{2}v_{2}+O(\eps^{3}).\nonumber
\end{eqnarray}
In this case, the turning functions $\lambda^{\pm}[u_{\eps},v_{\eps}]$ can be approximated by 
\begin{align*}
\lambda^{\pm}[u_{\eps},v_{\eps}]\approx &\lambda_{1}+ \lambda_{2} \lambda_3 f(K^N \ast u_0) + \lambda_{3}f\big(y_{D}^{\pm}[u_{0},v_{0}]\big)+O(\epsilon).
\end{align*}
If we impose the initial conditions 
\begin{equation}
v_{0}=\frac{\gamma \lambda_{3}\Big(f\big(y_{D}^{-}[u_{0},v_{0}]\big)-f\big(y_{D}^{+}[u_{0},v_{0}]\big) \Big)u_0}{2\lambda_{1}+2\lambda_{2} \lambda_3  f(y_{N}[u_0]) + \lambda_{3}\Big(f\big(y_{D}^{-}[u_{0}, v_{0}]\big)+f\big(y_{D}^{+}[u_{0},v_{0}]\big) \Big)},\label{IC}
\end{equation}
then, at $O(\eps^{0})$, equation (\ref{Cattaneo_epsilon;a}) reduces to
\begin{equation}
\frac{\partial u_{0}}{\partial t}+\frac{\partial}{\partial x}\big(u_{0}F[u_{0},v_{0}] \big)=0,\label{eq_hyp_F}
\end{equation}
with 
\begin{equation}
F[u_{0},v_{0}]=\frac{\gamma \lambda_{3}\Big(f\big(y_{D}^{-}[u_{0}, v_{0}]\big)-f\big(y_{D}^{+}[u_{0},v_{0}]\big)\Big)}{2\lambda_{1}+2\lambda_{2} \lambda_3 f(y_N[u_0]) + \lambda_{3}\Big(f\big(y_{D}^{-}[u_{0},v_{0}]\big)+f\big(y_{D}^{+}[u_{0},v_{0}]\big) \Big)}.\label{Eq_F}
\end{equation}

Note that for model M2 (see equations (\ref{1D_model_M2y})+(\ref{1D_model_M2yN})), in the absence of alignment (i.e., when $q_{al}=0$), there is no $v_{0}$ term.
Moreover, if we assume that the turning functions are linear, i.e., $f(y^{\pm}_{D}[u,v])=\pm y_{D}[u,v]$, then $f(y_{D}^{+}[u,v])+f(y_{D}^{-}[u,v])=0$, $f(y_{D}^{+}[u,v])-f(y_{D}^{-}[u,v])=2y_{D}[u,v]$, and thus the non-local directional term in the denominator of (\ref{Eq_F}) vanishes. Thus, in the case $q_{al}=0$ and $\lambda_{2} = 0$, equation (\ref{eq_hyp_F}) reduces to
\begin{equation}
 \frac{\partial u}{\partial t}+\frac{\partial}{\partial x}\big(uV[u] \big)=0,\label{eq_hyp_1d}
 \end{equation}
 with $V[u]= - c K\ast u$, $c=\gamma \lambda_{3}/\lambda_{1}$. Equation (\ref{eq_hyp_1d}) is commonly used to describe self-organised aggregations that result from attractive-repulsive interactions \cite{Topaz:Bertozzi,HackettFellner_NonlocalKinetic, Topaz:Leverentz,CHM2014}. In all these models, the usual choice for interaction kernels are Morse potentials (Fig. \ref{Spatial_ranges}(b)). A particularly interesting case is presented in \cite{MillerKolpas_3Zone}, where the authors incorporate also alignment interactions into the velocity function $V[u]$:
 \begin{equation*}
 V[u]=K_{r}\ast u+G_{al}\ast (uV[u])(G_{al}\ast u)^{-1}+K_{a}\ast u.
 \end{equation*}
 The orientation term $G_{al}\ast (uV[u])(G_{al}\ast u)^{-1}$ is defined implicitly in terms of the velocity $V$, by assuming the velocity inside the orientation/alignment zone depends on the average velocity of individuals. (This assumption leads to non-unique velocity solutions.) In this model, the kernels are all Gaussian (see Fig. \ref{Spatial_ranges}(c)). Returning now to equation (\ref{eq_hyp_F}), we note that in the presence of alignment interactions ($q_{al}>0$), the velocity depends on the flux $v_{0}$ of individuals, which is defined implicitly through the initial conditions in (\ref{IC}).  
 
\section{Description of 2D Models}\label{model_description}

An attempt to generalise the 1D model (\ref{1D_model})-(\ref{1D_model_M2y})-(\ref{1D_model_M2yN}) to two dimensions was made by Fetecau \cite{Fetecau2010}. The Boltzman-type model described in \cite{Fetecau2010} incorporates the non-local social interactions in the reorientation terms:
\begin{equation}
\frac{\partial u}{\partial t}+\gamma \v{e}_{\phi}\cdot \nabla_{\mathbf{x}}u=-\lambda(\mathbf{x},\phi)u+\int_{-\pi}^{\pi}T(\mathbf{x},\phi',\phi)u(\v{x},\phi',t)d\phi'.\label{Model_RF}
\end{equation}
Here, $u(\mathbf{x},\phi,t)$ is the total population density of individuals located at $\mathbf{x}=(x,y)$, moving at a constant speed $\gamma>0$ in direction $\phi$. The term $\v{e}_{\phi}=(\cos(\phi),\sin(\phi))$ gives the movement direction of individuals. The reorientation terms, $\lambda(\mathbf{x},\phi)$ and $T(\mathbf{x}, \phi',\phi)$ depend on the non-local interactions with neighbours, which can be positioned in the repulsive, attractive, and alignment ranges depicted in Fig. \ref{Spatial_ranges}(a). Thus, these terms have three components each, corresponding to the three social interactions:
\begin{eqnarray*}
T(\mathbf{x},\phi',\phi)=T_{al}(\mathbf{x},\phi',\phi)+T_{a}(\mathbf{x},\phi',\phi)+T_{r}(\mathbf{x},\phi',\phi).\label{T2D}
\end{eqnarray*}
In contrast to the model in \cite{Fetecau2010}, here we assume that the reorientation terms $\lambda_j (\v{x}, \phi') = \int T_j(\mathbf{x},\phi',\phi) \rd \phi$, $j = r, a, al$ have both a constant and a density-dependent component:
\begin{align}
T&_{al}(\mathbf{x},\phi',\phi)= \frac{\eta_{al}}{2 \pi}+\,\label{T_al}\\ 
 &  \lambda_3 \, q_{al}\int_{-\pi}^{\pi}\int_{\mathbb{R}^{2}}K_{al}^{d}(\mathbf{x}-\mathbf{s})K_{al}^{o}(\theta,\phi')\omega_{al}(\phi'-\phi,\phi'-\theta)u(\mathbf{s},\theta,t)ds d\theta,\notag\\
T&_{r,a}(\mathbf{x},\phi',\phi)= \frac{\eta_{r,a}}{2 \pi}+ \,   \label{T_ra}\\
&\lambda_3 \,q_{r,a}\int_{-\pi}^{\pi} \int_{\mathbb{R}^{2}} K_{r,a}^{d}(\mathbf{x}-\mathbf{s})K_{r,a}^{o}(\mathbf{s},\mathbf{x},\phi')\omega_{r,a}(\phi'-\phi,\phi'-\psi)u(\mathbf{s},\theta,t)ds d\theta.\notag
\end{align}
\begin{remn} 
By defining the constant basic turning rate to be $\lambda_1 = \eta_{r} + \eta_{al} + \eta_{a}$, we generalised the model in \cite{Fetecau2010} (where $\lambda_{1}=0$). Note that the turning rates here are linear functions of the non-local interactions with neighbours. This is in contrast to the more general non-linear turning function $f$ we considered in Section \ref{Parabolic_limit} for the 1D hyperbolic model. 

In what follows, we are interested in nonconstant turning rates $\lambda_j(\v{x}, \phi')$, $j=r,a,al$, and so we will henceforth assume $\lambda_3 \neq 0$.
\end{remn}

As in \cite{Fetecau2010}, $\lambda_j$, $j=r,a,al$, are defined in terms of both distance kernels and orientation kernels. The 2D distance kernels $K_{j}^{d}$, $j=r,a,al$ are given by
\begin{equation}\label{distancekernels}
K_{j}^{d}(\mathbf{x})=\frac{1}{A_{j}}e^{-(\sqrt{x^2+y^2}-d_{j})/m_{j}^{2}}, \;\; j=r,a,al,
\end{equation}
where constants $A_{j}$ are chosen such that the kernels integrate to one. The orientation kernels $K_{j}^{o}$ measure the likelihood of turning in response to the movement direction of neighbours (for alignment interactions) or in response to the position of neighbours (for repulsive and attractive interactions):
\begin{eqnarray}
 K_{al}^{o}(\theta,\phi)&=&\frac{1}{2\pi}(1-\cos(\phi-\theta)),\nonumber\\
 K_{r,a}^{o}(\mathbf{s},\mathbf{x},\phi)&=&\frac{1}{2\pi}(1\pm \cos(\phi-\psi)),\nonumber
\end{eqnarray} 
where $\psi$ is the angle between the positive $x$-axis and the relative location $\mathbf{s}-\mathbf{x}$ of the neighbours at $\mathbf{s}$ with respect to the reference individual at $\mathbf{x}$.
Finally, $\omega$ describes the tendency to turn from direction $\phi'$ to direction $\phi$, as a result of interactions with individuals moving in direction $\theta$:
\begin{equation*}
\omega(\phi'-\phi,\phi'-\theta)=g(\phi'-\phi-R(\phi'-\theta)), 
\end{equation*} 
for some suitable choice of $g$. Note that in the case $\lambda_1 = 0$, the function $\omega$ describes the probability of re-orientation and thus we require $\int \omega(\phi'-\phi,\phi'-\theta) d\phi = 1$. For example, $g$ could be a periodic function that integrates to one:
\begin{equation*}
g(\theta) = \frac{1}{\sqrt{\pi}\sigma}\sum_{z\in\mathbb{Z}}e^{-(\frac{\theta+2\pi z}{\sigma})^{2}}, \;\; \theta\in(-\pi,\pi), \label{funct_g}
\end{equation*}
with $\sigma$ a parameter measuring the uncertainty of turning (with small $\sigma$ leading to exact turning) \cite{Fetecau2010,GeigantLadizhanskyMogilner}. Another typical choice could be the Von-Misses distribution, as in Vicsek-type models \cite{DegondVicsek-vonMisses}. 
On the other hand, when $\lambda_1>0$, then $\omega$ satisfies $\int \omega(\phi'-\phi,\phi'-\theta) d\phi = 0$ and therefore $g$ is required to be odd. 

For simplicity, the 2D kinetic model (\ref{Model_RF}) can be re-written as
\begin{equation*}
\frac{\partial u}{\partial t}+\gamma e_{\phi}\nabla_{x}u=-Q^{-}[u]+Q^{+}[u,u], \label{Kinetic_simplified}
\end{equation*}
with 
\begin{subequations}
\begin{align}
Q^{-}[u]&=Q^{-}_{r}[u]+Q^{-}_{a}[u]+Q^{-}_{al}[u],\nonumber\\
Q^{-}_{j}[u]&=\lambda_{j}(x,\phi) u, \;\; j=r,al,a,\nonumber\\
Q^{+}[u,u]&=Q_{r}^{+}[u,u]+Q_{a}^{+}[u,u]+Q_{al}^{+}[u,u],\nonumber\\
Q^{+}_{j}[u,u]&=\int_{-\pi}^{\pi}T_{j}(x,\phi',\phi)u(x,\phi',t)d\phi',\;\; j=r,al,a.\nonumber
\end{align}
\end{subequations}

\begin{remn}
Fetecau \cite{Fetecau2010} showed that by imposing the turning angle to have only two possible values $\phi=\pm \pi$, the 2D model (\ref{Model_RF}) can be reduced to the 1D model (\ref{1D_model}) for some turning rates $\lambda^\pm[u^+,u^-]$. Considering the more general turning rates (\ref{T_al}) and (\ref{T_ra}), we recover (\ref{eq_lambda}) with $\lambda_1,\lambda_{3} \geq 0$, $\lambda_2=0$ for a linear turning function $f(z)=z$, and with the communication mechanism
\begin{align*}
 y^{\pm}_D[u^+,u^-]= &\,\frac{1}{\pi} q_{al} \int_{-\infty}^{\infty} K_{al}(x-s)\left( u^{\mp}(s,t) \right) \rd s\\
  &+\frac{1}{\pi} q_a \int_{-\infty}^x K_a(x-s)\left(u^+(s,t) + u^-(s,t)\right) \rd s\\
  &+\frac{1}{\pi} q_r \int^{\infty}_x K_r(x-s) \left(u^+(s,t) + u^-(s,t)\right)  \rd s.
\end{align*}
This is a similar turning behaviour to model M2 in \cite{Eftimie2}, since individuals receive and emit omni-directional communication signals, but with the function $f$ linear.
\end{remn}

The diffusion limit (i.e., $x=x^*/\eps$, $t=t^{*}/\eps^{2}$) of a transport model similar to (\ref{Model_RF}), but with constant turning rates $\lambda$ was discussed in \cite{Hillen:Othmer, Othmer:Hillen}. In the following we consider the parabolic limit for model (\ref{Model_RF}) with density-dependent turning rates.

\subsection{Parabolic Drift-Diffusion limit}\label{Sect:ParabLimit2D}

We focus on the case where individuals are only influenced slightly by the presence of neighbours, i.e., the turning mechanism can be assumed to be a small perturbation of a uniform turning probability. 
In this case, we will show that the Boltzmann-type equation (\ref{Model_RF}) can be reduced to a drift-diffusion equation in the macroscopic regime.

We consider the scaling $t=t^*/\eps^{2}, \v{x}=\v{x}^*/\eps$, where $\eps \ll 1$ is a small parameter, and assume the turning rate to be the result of (i) a constant random turning rate ($\lambda_1 = \eta_{al} + \eta_r + \eta_a$), (ii) a random density-dependent turning rate ($\lambda_{2}$) that describes individuals randomly turning towards neighbours, every time they perceive them, and (iii) a very weak density-dependent turning rate ($\epsilon \lambda_{3}$) that describes individuals turning towards/away from neighbours, or aligning with their neighbour's movement direction. The latter turning rate is the result of time and space rescaling, which leads to a reduced perception of directionality of neighbours. Thus,
\begin{equation} \label{Teps}
 T[u](\v{x}, \phi', \phi) = \frac{\lambda_1}{2 \pi} + \lambda_3 \left(\frac{\lambda_2}{2 \pi} \, K^d \ast \rho(\v{x}, t) + \eps  \,  B[u](\v{x}, \phi', \phi)\right),
\end{equation}
with $ \rho(\v{x}, t) = \int_{-\pi}^\pi u(\v{x}, \phi, t) \, \rd \phi$, and where we define $K^d(\v{x}):= q_{al} K^d_{al}(\v{x}) + q_{a} K^d_{a}(\v{x}) + q_{r} K^d_{r}(\v{x})$ to be the \emph{social distance kernel}.
If $\lambda_3 \neq 0$, the \emph{social response function} $B[u]$ can be derived from assumptions on the re-orientation function $\omega$ and the orientation kernels $K_{j}^{o}$,
\begin{align*}
g_j (\vartheta) &= \lambda_2+ \eps  G_j (\vartheta),\\
 K^{o}_{al}(\theta, \phi) &= \frac{1}{2 \pi} \left( 1 - \eps \, \cos \left( \phi - \theta \right)\right),\\
  K^{o}_{r,a}(\v{s}, \v{x}, \phi) &= \frac{1}{2 \pi} \left( 1 \pm \eps \, \cos \left( \phi - \psi \right)\right),
\end{align*}
where $G_j(\vartheta)$, $j=r, a, al$ are \emph{signal response functions} to be chosen according to the biological context. These assumptions reflect the hypothesis that an individual's turning behaviour is only influenced slightly by the presence of neighbours. If $\lambda_1=0$, we further have $\int_{-\pi}^{\pi} G_j (\phi' - \phi - R(\phi' - \theta)) \rd \phi = 0$, $j=r, a, al$ as the probability to turn to any new angle is 1. In addition, we want the turning function $R(\vartheta)$ to be close to an unbiased turning mechanism. This can be expressed by taking $R(\vartheta) = \eps \vartheta$, which indeed corresponds to weak interaction between individuals, \cite{GeigantLadizhanskyMogilner}. We obtain $B[u] = B_{al}[u] + B_a[u] + B_r [u]$ with
\begin{align}
 B_{al}[u](\phi', \phi)= &\frac{1}{2 \pi}\,q_{al} \, G_{al}(\phi' - \phi)\,K_{al}^d \ast 					\rho(\v{x}, t) \label{Bal}\\
	      &-\frac{\lambda_2}{2 \pi } \, q_{al} 
	      \int_{\R^2} K^d_{al}(\v{x} - \v{s}) \int_{-\pi}^\pi \cos\left( \phi' - \theta \right) \, u(\v{s}, \theta, t)\, \rd \theta \, \rd \v{s}, \notag\\
  B_{r,a}[u](\phi', \phi)= &\frac{1}{2 \pi}\,q_{r,a} \, G_{r,a}(\phi' - \phi)\,K_{r,a}^d \ast 					\rho(\v{x}, t) \label{Bra}\\
	      &\pm \frac{\lambda_2}{2 \pi} \, q_{r,a} 
	      \int_{\R^2} K^d_{r,a}(\v{x} - \v{s}) \cos\left( \phi' - \psi \right) \, \rho(\v{s}, t) \, \rd \v{s}.\notag
\end{align}
Let us introduce
\begin{align*}
 K_*^d (\v{x}^*) = \frac{1}{\eps} K^d\left( \frac{\v{x}^*}{\eps}\right), \quad
B_*(\v{x}^*,\phi', \phi) = \frac{1}{2 \pi}  B\left(\frac{\v{x}^*}{\eps},\phi', \phi\right).
\end{align*}
Simplifying the notation by dropping $*$, system (\ref{Model_RF}) writes in the new variables as
\begin{align} \label{scaledmodel}
                \, \eps^{2}\partial_{t} u + \eps \, \v{e}_{\phi} \cdot  \nabla_{\v{x}} u = \, 
                &  \frac{1}{2 \pi}\left(\lambda_1 +  \lambda_2\, \lambda_3 \, K^d \ast \rho  \right)
                \left( \rho - 2\pi u \right) \\ \notag
                &+ \eps \, \lambda_3 \, 2 \pi \, \left.
                \int_{-\pi}^\pi B(\v{x},\phi', \phi) \, u(\v{x},\phi', t) \,\rd \phi'\right. \\ \notag
                & - \eps \,\lambda_3 \, 2 \pi \,  \left. u(\v{x},\phi, t) \, \int_{-\pi}^\pi B (\v{x},\phi, \phi') \rd \phi'
                \right. .        
            \end{align}
Using a Hilbert expansion approach, $u = u_0 + \eps u_1 + \eps^2 u_2 + ...$, and defining the macroscopic densities $\rho_i = \int_{-\pi}^\pi u_i \, \rd \phi$ for $i \in \N_0$, we obtain at leading oder a relaxation towards a uniform angular distribution at each position:
\begin{align} \label{order-2a}
  u_0(\v{x}, \phi, t)  &= \rho_0(\v{x}, t) F(\phi),\\
  F(\phi) &= \frac{1}{2\pi} \mathbbm{1}_{\phi \in (-\pi, \pi]}. \notag
\end{align}
Integrating (\ref{scaledmodel}) with respect to the direction of motion $\phi$, we obtain the continuity equation
\begin{equation}\label{4}
 \partial_t \rho_0 + \int_{-\pi}^\pi \v{e}_{\phi} \cdot  \nabla_{\v{x}} \, u_1 \rd \phi = 0.
\end{equation}
Comparing orders of $\eps$, we can derive an expression for $u_1$ in terms of $u_0, \rho_0, \rho_1$. 
Substituting into (\ref{4}), we arrive at a macroscopic drift-diffusion equation of the form
\begin{equation*}
 \partial_t \rho_0 = 
  \, \nabla_{\v{x}} \, . \, \left(D[\rho_0]\,  \nabla_{\v{x}} \rho_0
 - \rho_0 \v{k}[\rho_0]
 \right),
\end{equation*}
where the macroscopic diffusion coefficient $D[\rho_0] = \gamma^2/(2(\lambda_1 + \lambda_2\, \lambda_3\, K^d \ast \rho_0 ))$ and the social flux
\begin{equation}\label{kwithB}
 \v{k}[\rho_0] =  \frac{\lambda_3\,\gamma}{\lambda_1 + \lambda_2\, \lambda_3 \,  K^d \ast \rho_0} \,
 \int_{-\pi}^\pi \int_{-\pi}^\pi \left(\v{e}_{\phi} - \v{e}_{\phi'}\right) 
  B[\rho_0](\v{x},\phi', \phi) \rd \phi' \rd \phi
\end{equation}
are both described in terms of microscopic quantities. 
In the context of collective behaviour of animal groups, we make two further assumptions:
\begin{enumerate}[(i)]
 \item Individuals can process information in a similar manner for all three types of social interactions:
 $$G_{al}(\vartheta) = G_{r}(\vartheta)= G_{a}(\vartheta)=: G(\vartheta) \quad \forall \vartheta. $$ 
 \item Individuals have symmetric perception, in other words, they can process information equally well from left and right. Then the turning probability function $\omega$ is bisymmetric, 
\begin{equation}
\omega(- \alpha, - \beta) = \omega(\alpha, \beta),\nonumber
\end{equation}
which implies symmetry of the signal response function $G$. 
\end{enumerate}
Under these assumptions, the first term of the social response functions $B_j[u]$ in (\ref{Bal}) and (\ref{Bra}) cancel when substituted into the social flux (\ref{kwithB}). Using (\ref{order-2a}), we can simplify the social flux even further and obtain the drift-diffusion equation
\begin{subequations} \label{macromodel}
\begin{align} 
 \partial_t \rho& = 
  \, \nabla_{\v{x}} \, . \, \left( D_{0}[\rho]\nabla_{\v{x}} \rho  \right)
 - \nabla_{\v{x}} \, . \, \left( \rho \,\v{k}[\rho] \right),\label{macromodel;a}\\
  D_{0}[\rho] &= \frac{\gamma^2}{2(\lambda_1 + \lambda_2\, \lambda_3\, K^d \ast \rho )}, \label{macromodel;b}\\
 \v{k}[\rho] (\v{x}, t)&=\frac{\lambda_2\, \lambda_3 \pi \gamma}{\lambda_1 + \lambda_2\, \lambda_3 \,K^d \ast \rho} 
 \, 
 \left(q_r\, K^d_r(\v{x}) \, \frac{\v{x}}{|\v{x}|} - q_a\, K^d_a(\v{x}) \, \frac{\v{x}}{|\v{x}|} \right) \ast \rho. \label{macromodel;c}
\end{align}
\end{subequations}
For notational convenience, we dropped the zero in $\rho_0$. Note that this equation is similar to the 1D drift-diffusion equation (\ref{Eq_parabolic_lambda2}) obtained via the parabolic limit for linear social interactions.

\begin{remn}
The turning rates $\lambda_1, \lambda_2, \lambda_3$ affect the interpretation of the turning function $g$ that appears in the expression of $T[u]$. In particular, if $\lambda_{1}\neq 0$, then $g$ can be interpreted as a small reorientation perturbation from the random turning behaviour. In this case, $g$ has to integrate to zero. On the other hand, if $\lambda_1=0$ (and $\lambda_2\, \lambda_3 = 1/(2\pi)$), then $g$ can be understood as a re-orientation probability in the sense discussed in \cite{GeigantLadizhanskyMogilner}. In this case, $g$ integrates to one. 
\end{remn}

\begin{remn}
In the case $\lambda_2=0$, the two-dimensional model (\ref{macromodel}) reduces to the heat equation, which is not the case in the parabolic limit (\ref{Eq_parabolic_lambda2}) of the corresponding one-dimensional hyperbolic model (\ref{1D_model}) with the turning rates given by (\ref{eq_lambda}). The reason for this is that although in both cases we assume linear interactions between individuals, $f(z)=z$, the 1D turning mechanism $y^\pm_D[u^+, u^-]$ defined in (\ref{eq_y}) does not depend on the density-dependent turning rate $\lambda_2$ in general (this is the case for example for models M2 (\ref{1D_model_M2y}) and M4 (\ref{1D_model_M4y}) we considered in Section 2). However, under scaling assumption (\ref{Teps}), the two-dimensional turning rate $$\lambda(\v{x}, \phi') = \lambda_1 + \lambda_2 \, \lambda_3 \, K^d \ast \rho(\v{x}, t) + \eps \, \lambda_3 \, \int B[u](\v{x}, \phi', \phi) \rd \phi$$
is of the form (\ref{eq_lambda}) with the two-dimensional turning function $ \int B[u]\rd \phi$ depending on $\lambda_2$. More precisely, the two-dimensional turning rate $\lambda(\v{x}, \phi')$ corresponds to (\ref{eq_lambda}) on the projected velocity set $\{0,\pi\}$, with a linear turning function $f(z)=z$ and with the non-directed and directed communication mechanisms given by 
\begin{align}
y_N^\pm[u]
 = &K^d\ast \rho (\v{x}, t),\notag\\
 y_D^\pm[u^+,u^-]
 = &\frac{G(0)+G(\pi)}{2} K^d\ast \rho (\v{x}, t)\label{2D_model_reduced}\\
  &\mp \lambda_2 \int_{\R} q_{al} K_{al}^d(\v{x}-\v{s}) \left( u^+(s_1, t) - u^-(s_1,t) \right) \rd s_1 \notag\\
  &\mp \lambda_2 \int_{-\infty}^{x_1} \left(q_{r} K_{r}^d(\v{x}-\v{s}) - q_{a} K_{a}^d(\v{x}-\v{s}) \right)\rho(\v{s},t) \rd s_1\notag\\
  &\pm \lambda_2 \int^{\infty}_{x_1} \left(q_{r} K_{r}^d(\v{x}-\v{s}) - q_{a} K_{a}^d(\v{x}-\v{s}) \right) \rho(\v{s},t) \rd s_1,\notag
\end{align}
where $\v{x}=(x_1,0)$, $\rho(\v{x},t)=u^+(x_1,t) + u^-(x_1,t) = u(x_1,t)$, and where we used assumptions (i) and (ii). Hence, model (\ref{1D_model}) with communication mechanism (\ref{2D_model_reduced}) yields a zero drift for $\lambda_2=0$ in the parabolic limit, and so (\ref{Eq_parabolic}) reduces likewise to the heat equation as expected.
\end{remn}

\begin{remn}
For some particular choices of distance kernels, the limiting parabolic model (\ref{macromodel}) can be reduced to well known equations. Let us assume, for example, that the distance kernels are constant on the whole domain,
\begin{equation}\label{constantkernels}
 K_j^d(\v{x}) = 1, \quad j = al, a, r.
\end{equation}
This assumption corresponds to a setting in which individuals interact equally well with all other individuals present in the entire domain. This is true locally for example if we have many individuals packed in little space. Under assumption (\ref{constantkernels}) together with $\lambda_1 = 0$ and $\lambda_3 = 1$, model (\ref{macromodel}) simplifies to

\begin{equation*}
 \partial_t \rho = 
  C_0 \, \, \Delta \rho + C_1 \, \nabla \, . \, \left( \rho \, \int_{\R^2} \v{e}_{\psi} \rho(\v{s})  \, \rd \v{s} \right),
\end{equation*}
where 
$$
\v{e}_\psi = \frac{\v{s} - \v{x}}{|\v{s} - \v{x} |},
$$
and $C_0, C_1$ are constants depending only on $\gamma, q_{al}, q_a, q_r$ and the total mass $\int \,\rho \, \rd \v{x}$.  
If $q_a = q_r$, then the attraction and repulsion forces cancel out and we obtain the heat equation. Let us henceforth assume $q_a \neq q_r$. Furthermore, we can write the social flux as
\begin{equation}\label{socialfluxpotential} 
 \v{k}[\rho] = \nabla W \ast \rho,
\end{equation}
where the interaction potential $W: \R^2 \longrightarrow \R$ is given by $ W(\v{x}) = C_1 |\v{x}|$.
In fact, for the more general distance kernels (\ref{distancekernels}) the social flux can also be written in the form (\ref{socialfluxpotential}), with the interaction potential $W$ behaving like $|\v{x}|$ close to zero and decaying exponentially fast as $|\v{x}| \longrightarrow \infty$ (e.g., Morse potentials). 
Therefore, we recover the diffusive aggregation equation 
\begin{align*}
 \partial_t \rho = 
  \Delta \rho + \nabla \, . \, \left( \rho \, \left( \nabla W \ast \rho \right) \right),
\end{align*}
which models the behaviour of particles interacting through a pairwise potential while diffusing with Brownian motion. This type of equation has received a lot of attention in recent years because of its ubiquity in modelling aggregation processes, such as collective behaviour of animals \cite{M&K, Morale,BertozziCarrilloLaurent,ChuangDOrsogna07} and bacterial chemotaxis \cite{Blanchet06} (see also the references therein). 
\end{remn}

\subsection{Grazing collision limit}

In the following, we focus on the case where individuals turn only a small angle upon interactions with neighbours. Note that many migratory birds usually follow favourable winds or magnetic fields \cite{Newton_MigratoryBirds}, and thus social interactions with neighbours, while useful to maintain the direction of the whole flock, might not have a considerable impact on directional changes of individual birds. Hence, the assumption of small turning angles following inter-individual interactions could be biologically realistic. This assumption corresponds to the so-called \textit{grazing collisions}, i.e., collisions with small deviation. In this case, we will show that the Boltzmann-type equation (\ref{Model_RF}) can be reduced to a Fokker-Planck equation with non-local advective and diffusive terms in the orientation space.

To keep the analysis simpler, let us focus for now only on the alignment interactions (i.e., assume $q_{a}=q_{r}=0$, and hence $Q^{\pm}_{j}=0$, $j=r,a$). The analysis of attraction and repulsion interactions is similar. The grazing collision assumption suggests that we can rescale the probability of re-orientation as  follows:
\begin{equation*}
\omega_{al}^{\eps}\left( \phi - \phi', \phi - \theta \right)=\frac{1}{\eps}g_{\eps}\Big(\frac{\phi-\phi'-\eps R(\phi-\theta)}{\eps} \Big).
\end{equation*}
Here, the parameter $\epsilon$ is related to the small re-orientation angle following interactions with neighbours moving in direction $\theta$. If we denote by $\eps \beta=\phi-\phi'-\eps R(\phi-\theta)$, then since $\omega_{\eps}$ integrates to 1, we obtain:
\begin{eqnarray}
1=\int_{-\pi}^{\pi}\omega_{al}^{\eps}(\phi-\phi',\phi-\theta)d\phi'=\int_{-\pi+\phi-R(\phi-\theta)}^{\pi+\phi-R(\phi-\theta)}g_{\eps}(\beta)d\beta=\int_{-\pi}^{\pi}g_{\eps}(\beta)d\beta, \nonumber
\end{eqnarray}
by periodicity of $g_{\eps}$.

Generally, when an interaction kernel in the Boltzmann equation presents a singularity point, the troubles are avoided by considering a weak formulation \cite{Goudon_BoltzmannFokkerPlank, Carillo09_Kinetic_Attraction-Repulsion}:
for all $\psi \in C_c^\infty([-\pi, \pi])$,
\begin{equation}
\int_{-\pi}^{\pi}\frac{\partial u}{\partial t}\psi(\phi)d\phi+\int_{-\pi}^{\pi}\gamma e_{\phi}\nabla_{x}u\psi(\phi)d\phi=\int_{-\pi}^{\pi}(-Q^{-}_{al}[u]+Q^{+}_{al}[u,u])\psi(\phi)d\phi. \label{2D_WeakForm}
\end{equation}
Then, expanding the RHS of (\ref{2D_WeakForm}) and defining $Q_{al}[u]= - Q_{al}^{-}[u]+Q^{+}_{al}[u,u]$, we obtain
\begin{align}
\int_{-\pi}^{\pi}Q_{al}[u]\psi(\phi)d\phi=
\eta_{al} \int_{-\pi}^{\pi} &\left( \frac{1}{2\pi}\rho(x,t) - u(x, \phi, t) \right) \psi(\phi) d\phi \nonumber\\
+\int_{-\pi}^{\pi}\int_{-\pi}^{\pi}\int_{\mathbb{R}^{2}}&\lambda_3 q_{al}K_{al}^{d}(x-s)K_{al}^{0}(\theta,\phi)u(x,\phi,t)u(s,\theta,t)\cdot \nonumber \\
\int_{-\pi}^{\pi}\omega_{al}^{\eps}&(\phi-\phi',\phi-\theta)\big[\psi(\phi')-\psi(\phi) \big]d\phi' ds d\theta d\phi. \label{RHS}
\end{align}
By substituting $\phi'=\phi-\eps \beta-\eps R(\phi-\theta)$ into the $\psi(\phi')$ term in (\ref{RHS}), and then expanding in Taylor series about $\phi$ we obtain:
\begin{align*}
\int_{-\pi}^{\pi}\omega_{al}^{\eps}&(\phi-\phi',\phi-\theta)\Big[\psi(\phi')-\psi(\phi) \Big]d\phi'\approx \\ &\int_{-\pi}^{\pi} g_{\eps}(\beta)\Big[\big(-\eps \beta -\eps R(\phi-\theta)\big)\frac{\partial \psi}{\partial \phi}  + 
\frac{\eps^{2}}{2}\big(\beta+R(\phi-\theta)\big)^{2}\frac{\partial^{2}\psi}{\partial\phi^{2}}\Big] d\beta.
\end{align*}
Equation (\ref{RHS}) can thus be approximated by

\begin{align}
\int_{-\pi}^{\pi}Q_{al}[u]\psi(\phi)d\phi=&\,
\eta_{al} \int_{-\pi}^{\pi} \left( \frac{1}{2\pi}\rho(x,t) - u(x, \phi, t) \right) \psi(\phi) d\phi\nonumber\\
&-\int_{-\pi}^{\pi}\frac{\partial }{\partial \phi}\Big[u(x,\phi,t)C_{al}^{\eps}[u,x,\phi]\Big]\psi(\phi)d\phi\nonumber\\
&+\int_{-\pi}^{\pi}\frac{\partial^{2}}{\partial \phi^{2}}\Big[u(x,\phi,t)D_{al}^{\eps}[u,x,\phi]\Big]\psi(\phi)d\phi,\nonumber
\end{align}
with the definitions

\begin{align*}
C_{al}^{\eps}[u,x,\phi]=&\int_{-\pi}^{\pi}\int_{\mathbb{R}^{2}}\lambda_3 q_{al}K_{al}^{d}(x-s)K_{al}^{0}(\theta,\phi)A_{al}^{\eps}(\phi-\theta)u(s,\theta,t)d\theta ds,\nonumber\\
D_{al}^{\eps}[u,x,\phi]=&\int_{-\pi}^{\pi}\int_{\mathbb{R}^{2}} \lambda_3 q_{al}K_{al}^{d}(x-s)K_{al}^{0}(\theta,\phi)B_{al}^{\eps}(\phi-\theta)u(s,\theta,t)d\theta ds,\nonumber
\end{align*}
where

\begin{align*}
A_{al}^{\eps}(\phi-\theta)=&-\eps \big(M_{1}(\eps)+ M_{0}(\eps)R(\phi-\theta)\big),\nonumber\\
B_{al}^{\eps}(\phi-\theta)=&\frac{\eps^{2}}{2}\big(M_{2}(\eps)+2M_{1}(\eps)R(\phi-\theta)+M_{0}(\eps)R(\phi-\theta)^{2} \big),\nonumber
\end{align*}
and $M_{n}(\eps)=\int_{-\pi}^{\pi}\beta^{n}g_{\eps}(\beta)d\beta$, $n=0,1,2,$ denote the moment generating functions of $g_{\eps}(\beta)$.
In a similar manner we can approximate the attractive and repulsive non-local terms:

\begin{align*}
\int_{-\pi}^{\pi}Q_{r,a}[u]\psi(\phi)d\phi=&\,\eta_{r,a} \int_{-\pi}^{\pi} \left( \frac{1}{2\pi}\rho(x,t) - u(x, \phi, t) \right) \psi(\phi) d\phi\nonumber\\
&-\int_{-\pi}^{\pi}\frac{\partial}{\partial \phi}\Big(u(x,\phi,t)C_{r,a}^{\eps}[u,x,\phi] \Big)\psi(\phi)d\phi\nonumber\\
&+\int_{-\pi}^{\pi}\frac{\partial^{2}}{\partial \phi ^{2}}\Big(u(x,\phi,t)D_{r,a}^{\eps}[u,x,\phi] \Big)\psi(\phi)d\phi,\nonumber
\end{align*}
where
\begin{align*}
C_{r,a}^{\eps}[u,x,\phi]=&\int_{-\pi}^{\pi}\int_{\mathbb{R}^{2}}\lambda_3 q_{r,a}K_{r,a}^{d}(x-s)K_{r,a}^{0}(s,x,\phi)A^{\eps}_{r,a}(s,x,\phi)u(s,\theta,t)ds d\theta,\nonumber\\
D_{r,a}^{\eps}[u,x,\phi]=&\int_{-\pi}^{\pi}\int_{\mathbb{R}^{2}}\lambda_3 q_{r,a}K_{r,a}^{d}(x-s)K_{r,a}^{0}(s,x,\phi)B_{r,a}^{\eps}(s,x,\phi)u(s,\theta,t)dsd\theta,\nonumber\\
A_{r,a}^{\eps}(s,x,\phi)=&-\eps(M_{1}(\eps)M_{0}(\eps)R(\phi-\psi_{s})),\nonumber\\
B_{r,a}^{\eps}(s,x,\phi)=&\frac{\eps^{2}}{2}\Big[M_{2}(\eps)+2M_{1}(\eps)R(\phi-\psi_{s})+M_{0}(\eps)R(\phi-\psi_{s})^{2} \Big].\nonumber
\end{align*}
Therefore, the kinetic model (\ref{Model_RF}) in the strong formulation can be approximated (when individuals turn only by a small angle upon interactions with their neighbours) by the following Fokker-Planck model:

\begin{align}
\frac{\partial u}{\partial t}+\gamma e_{\phi}\cdot \nabla_{x}u=
&\,\lambda_1 \left( \frac{1}{2\pi}\rho(x,t) - u(x, \phi, t) \right) \label{2D_FokkerPlanck}\\
&+ \, \frac{\partial}{\partial \phi}\Big[-uC^{\eps}[u,x,\phi]+\frac{\partial}{\partial \phi}(uD^{\eps}[u,x,\phi]) \Big],\nonumber
\end{align}
with $\lambda_{1}=\eta_{a}+\eta_{al}+\eta_{r}$ and
\begin{eqnarray*}
C^{\eps}[u,x,\phi]=C_{al}^{\eps}[u,x,\phi]+C_{r}^{\eps}[u,x,\phi]+C_{a}^{\eps}[u,x,\phi],\nonumber\\
D^{\eps}[u,x,\phi]=D_{al}^{\eps}[u,x,\phi]+D_{r}^{\eps}[u,x,\phi]+D_{a}^{\eps}[u,x,\phi].
\end{eqnarray*}
While non-local 2D Fokker-Planck models have been introduced in the past years in connection to self-organised aggregations, the majority of these models consider local diffusion \cite{DegondMotsch_OrientationLimitMacro, BarbaroDegond2013}. 
If we neglect the $\eps^{2}$ terms (i.e., $B^{\eps}\approx 0$) and assume $\lambda_1 = 0$, equation (\ref{2D_FokkerPlanck}) reduces to a Vlasov-type flocking equation:

\begin{equation*}
\frac{\partial u}{\partial t}+\gamma e_{\phi}\cdot \nabla_{x}u+\frac{\partial}{\partial \phi}\Big[uC^{\eps}[u,x,\phi] \Big]=0. \label{2D_Vlasov}
\end{equation*}
These type of models have been previously derived from individual-based models (Vicsek or Cucker-Smale models) with or without noise \cite{DegondMotsch_OrientationLimitMacro,HaTadmor_ParticleKineticHydrodynamic,Carillo09_Kinetic_Attraction-Repulsion}.

\section{Asymptotic Preserving Methods for 1D models}\label{Sect:asymptotic}

The kind of diffusion asymptotics we employed in the previous sections have been numerically investigated in \cite{CarrilloGoudon} using so-called asymptotic preserving (AP) schemes. AP methods, which improve the scheme already proposed in \cite{Godillon-Lafitte}, are a fully explicit variation of the methods introduced in \cite{Klar98, Klar99}. Taking advantage of our understanding of the limit process, we base our analysis on a splitting strategy with a convective-like step involving the transport part of the operator and an explicitly solvable ODE step containing stiff sources. 


\subsection{Odd and Even Parity}
  We consider the 1D kinetic model (\ref{1D_model}) written as an odd-even decomposition,
  \begin{equation*}
 \begin{cases}
  \partial_t r + \gamma \partial_x j &= 0,\\
  \partial_t j + \gamma \partial_x r &= -2 \lambda^+[r, j] (r + j) + 2 \lambda^-[r, j] (r-j),
 \end{cases}
\end{equation*}
  with the equilibrium part (macro part/even part) $r$ and the non-equilibrium part (micro part/odd part) $j$ given by
  $$
   r(x, t) = \frac{1}{2} \left( u^+(x, t) + u^-(x, t) \right), \quad
   j(x, t) = \frac{1}{2} \left( u^+(x, t) - u^-(x, t) \right).
  $$
 Under scaling assumption (\ref{fassumption}) for (\ref{eq_lambda}), this model reads in the new variables $x=\tilde x/\eps, t=\tilde t/\eps$ as follows:
  \begin{align*} 
  \eps \partial_{\tilde t} \tilde r +  \gamma \partial_{\tilde x} \tilde j = &0 \\
  \eps \partial_{\tilde t} \tilde j + \gamma \partial_{\tilde x} \tilde r 
  = &\, \tilde r \lambda_3 (f[\tilde y^-] - f[\tilde y^+])\notag\\
  &- \frac{1}{\eps} \tilde  j \left(2 \lambda_1 + 4 \eps \lambda_2 \lambda_3 f\left(\tilde{K}^N \ast \tilde r\right) + \eps \lambda_3 (f[\tilde y^+] + f[\tilde y^-])\right),\notag
\end{align*}
 where $\tilde{K}^N(\tilde x) = \frac{1}{\eps} K^N(\frac{\tilde x}{\eps})$. Rearranging the terms and dropping ``$\sim$" for notational convenience, we obtain for $r$ and $J := \frac{1}{\eps} \, j$:
\begin{align} \label{rJmicro}
 \begin{dcases}
  \partial_t r +  \gamma \partial_x J = & \hspace{-0.3cm}0\\
  \partial_t J + \gamma \partial_x r 
  = &
   \hspace{-0.3cm}\frac{1}{\eps^2} r \lambda_3 (f[ y^-] - f[ y^+]) + \left( 1 - \frac{1}{\eps^2} \right) \gamma \partial_x r\\
    & \hspace{-0.3cm}-\frac{1}{\eps^2} J\left(2 \lambda_1 + 4 \eps \lambda_2 \lambda_3 f\left(K^N \ast r\right) + \eps \lambda_3 (f[ y^+] + f[y^-]) \right).
   \end{dcases}
\end{align}

\subsection{Operator Splitting}
We can now employ an operator splitting method on (\ref{rJmicro}), separating the stiff source part, which can be treated by an implicit Euler method, and the transport part, which we can solve by an explicit method such as upwinding:
\begin{enumerate}
 \item \textbf{Stiff source part:}
\begin{align}\label{Operator1}
  \partial_t r\,  = &0, \notag\\
  \partial_t J 
  = &
  \frac{1}{\eps^2} r \lambda_3 (f[ y^-] - f[ y^+]) + \left( 1 - \frac{1}{\eps^2} \right) \gamma \partial_x r\\
  - &\frac{1}{\eps^2} J\left(2 \lambda_1 + 4 \, \eps \, \lambda_2 \lambda_3 f\left(K^N \ast r\right) + \eps \, \lambda_3 (f[ y^+] + f[y^-]) \right).\notag
\end{align}
\item \textbf{Transport part:}
\begin{align} \label{Operator2}
  \partial_t r +  \gamma \, \partial_x J &= 0,\\
  \partial_t J + \gamma \, \partial_x r \notag
  &= 0.
\end{align}
\end{enumerate}
It can easily be verified that, in the limit $\eps \to 0$, we recover indeed the macroscopic model (\ref{Eq_parabolic}) for $u=2r$.

\subsection{Alternated Upwind Discretisation}
 
 In the following, we are interested in the numerical implementation of model (\ref{1D_model}) with the turning rates (\ref{eq_lambda}) depending on a non-linear turning function $f$ without a non-directed density-dependent turning term (i.e. $\lambda_2=0$). As shown in Section \ref{Parabolic_limit}, in this case, the parabolic limit yields the drift-diffusion equation
 $$
 \partial_t u = D_0 \partial_{xx} u - B_0 \partial_x \left( u (f^-[u] - f^+[u]) \right),
 $$
 with $D_0=\gamma^2/(2 \lambda_1)$ and $B_0=\lambda_3 \gamma/(2 \lambda_1)$. Note the shortcut notation $f^\pm[u]=f(y_D^\pm[u])$. We propose an alternated upwind discretisation with the even part $r$ evaluated at full grid points $x_i = i \, \Delta x$, and the odd part $J$ evaluated at half grid points $x_{i+ \frac{1}{2}} = (i + \frac{1}{2}) \, \Delta x$. First, we discretise the stiff source part (\ref{Operator1}) using an implicit Euler discretisation and respecting the direction of the drift. We obtain an explicit expression for $J^*$,
 \begin{align*}
J^*_{i+\frac{1}{2}} \, = \, &\frac{
\eps^2 J^n_{i+\frac{1}{2}} + \gamma \frac{\Delta t}{\Delta x} \left(\eps^2 - 1 \right) \left(r_{i+1}^n - r_i^n \right)
}{
\eps^2 + 2 \lambda_1 \Delta t + \eps \lambda_3 \Delta t \left( f^+[r^n] + f^-[r^n] \right)_{i+\frac{1}{2}}
}\\
 &+
\frac{
  \lambda_3 \Delta t \left( 
 \left( f^-[r^n] - f^+[r^n] \right)^+_{i+\frac{1}{2}} \, r^n_i
  +\left(f^-[r^n] - f^+[r^n] \right)^{-}_{i+\frac{1}{2}} \, r^n_{i+1}
  \right)
}{
\eps^2 + 2 \lambda_1 \Delta t + \eps \lambda_3 \Delta t \left( f^+[r^n] + f^-[r^n] \right)_{i+\frac{1}{2}}
},
\end{align*}
with $r^* = r^n$. Here, $r^n$ and $J^n$ are the numerical solutions of $r$ and $J$ at time $t_n = n \Delta t$. We use the ``$*$"-notation for half steps in time. Since $J$ is evaluated at half grid point, the discretisation of the transport part (\ref{Operator2}) can be chosen independently of the sign of the drift,
\begin{align*} 
 \frac{1}{\Delta t} \left ( r_i^{n+1} -r_i^* \right)  + \frac{1}{\Delta x} \left( J_{i+\frac{1}{2}}^* - J_{i-\frac{1}{2}}^* \right) &= 0,\\
  \frac{1}{\Delta t} \left( J_{i+\frac{1}{2}}^{n+1} - J_{i+\frac{1}{2}}^* \right) + \frac{1}{\Delta x} \left( r_{i+1}^* - r_{i}^* \right) &= 0.\notag
\end{align*}
Taking the limit $\eps \to 0$ in the expression for $J^*_{i+\frac{1}{2}}$ and substituting into the first equation of the transport part, we obtain the following discretisation of the one-dimensional macroscopic model (\ref{macromodel}):
\begin{align*}
&\frac{u_i^{n+1} - u_i^n}{\Delta t} = \frac{D_0}{(\Delta x)^2} \left(\partial_{xx}^{(c)}u^n\right)_i\\
 &-\frac{B_0}{\Delta x}  \left(
 u_i^n \left( f^-[r^n] - f^+[r^n]\right)^+_{i+\frac{1}{2}} 
 - u_{i-1}^n \left( f^-[r^n] - f^+[r^n]\right)^+_{i-\frac{1}{2}} 
 \right)\\
 &-\frac{B_0}{\Delta x} \left(
 u_{i+1}^n \left( f^-[r^n] - f^+[r^n]\right)^-_{i+\frac{1}{2}} 
 - u_{i}^n \left(f^-[r^n] - f^+[r^n] \right)^-_{i-\frac{1}{2}} 
 \right).
 \end{align*}
 Here, $\partial_{xx}^{(c)} u^n$ denotes the standard central difference discretisations.
 This illustrates how the choice of discretisation for (\ref{Operator1}) directly induces a discretisation of model (\ref{macromodel}).\\
 
\begin{remn}
 The stability restriction for the proposed AP scheme is less clear. We can expect that the time steps size $\Delta t$ needs to be sufficiently small, with an upper stability bound depending on the space step size $\Delta x$, the diffusion coefficient $D_0$, and the social interaction kernels via the terms $K^N \ast u$ and $f^{\pm}[u]$. 
\end{remn}

 \subsection{Simulation results}\label{Sect:Sim}
 
In Section \ref{Sect:SS_stability} we have seen that for model M4, the two Hopf bifurcations that occurred for the $k_{4}$ and $k_{5}$ modes have disappeared as $\epsilon \to 0$. In this Section, we start with a rotating wave pattern (i.e., travelling pulses) that arises at $\epsilon=1$ through a Hopf bifurcation (i.e., for the same parameter values as in Figure \ref{Fig_Ss_M4_Epsilon}: $q_{a}=1.545$, $q_{r}=2.779$, $\lambda_{1}=0.2$, $\lambda_{3}=0.9$, $\gamma=0.1$, $A=2$). Then, we  investigate numerically what happens with this pattern as $\epsilon \to 0$. The initial conditions for the simulations are random perturbations -- of maximum amplitude $0.2$ -- of the spatially homogeneous steady state $u^{*}=A/2=1$. We start with $\epsilon=1$, and run the numerical simulations up to $t=1000$. Then we decrease $\epsilon$, and choose the new initial condition to be the final solution obtained with the previous $\epsilon$ value. 

\begin{figure}[!hp]
\centering
 \includegraphics[width=4.4in]{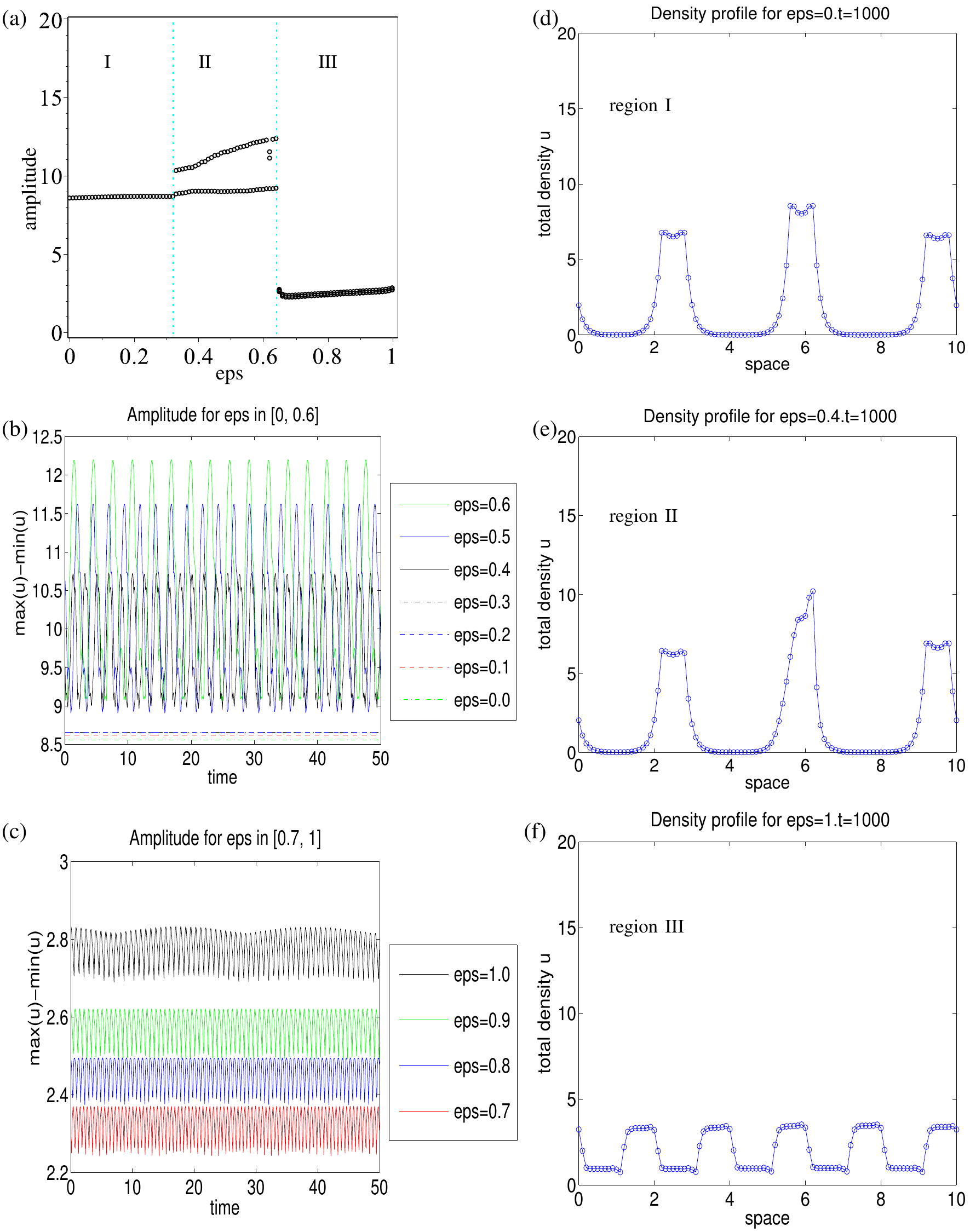}
\caption{The amplitude and density profile of the patterns obtained for $q_{a}=1.545$, $q_{r}=2.779$, $q_{al}=0$, $\lambda_{1}=0.2$, $\lambda_{3}=0.9$, as $\epsilon$ is decreased from $1.0$ to $0.0$.  (a) Bifurcation diagram for the amplitude of the patterns as a function of $\epsilon$. For $\epsilon\leq 0.32$ (region I), the amplitude is constant. For $\epsilon \in (0.32,0.64)$ (region II) the amplitude oscillates between two different values. For $\epsilon>0.64$ (region III) there are some very small oscillations in the amplitude, however due to the scale of the plot these oscillations are almost unobservable. (b) Amplitude of the patterns for $\epsilon\in(0,0.3)$ and  for $t\in(0,50)$. (c) Amplitude of the patterns for $\epsilon\in(0.5,1.0)$ and  for $t\in(0,50)$. (d) Density profile for the patterns observed in region I; (e) Density profile for the patterns observed in region II; (f) Density profile for the patterns observed in region III.}
\label{Fig_Amplitudes_Epsilon}
\end{figure}

\begin{figure}[!t]
\centering
 \includegraphics[width=5.8in]{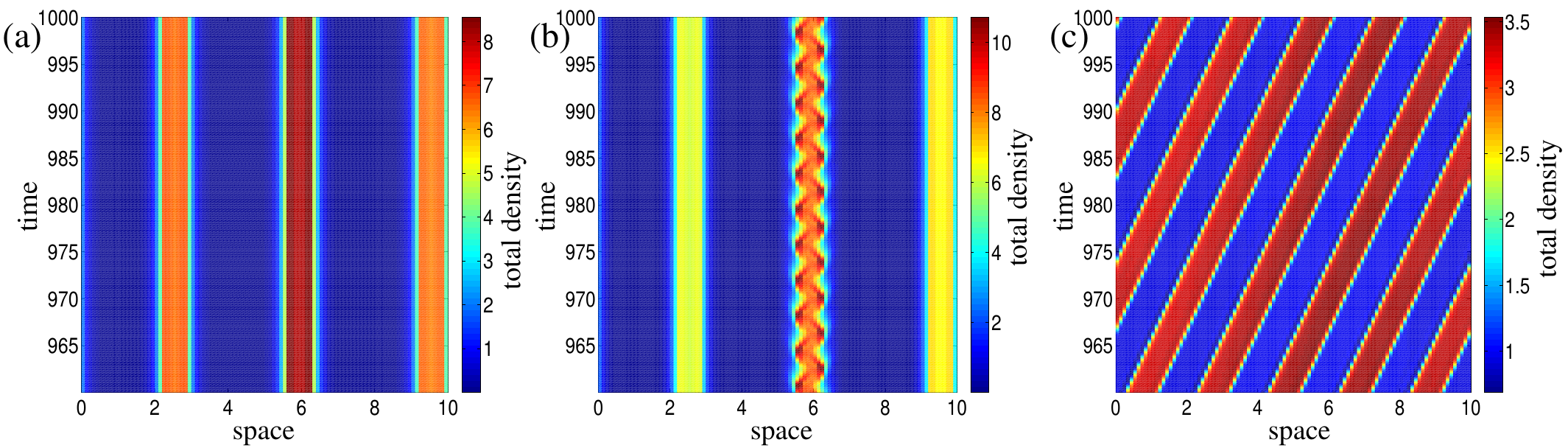}
\caption{The spatial and spatio-temporal patterns obtained with model M4, for $q_{a}=1.545$, $q_{r}=2.779$, $q_{al}=0$, $\lambda_{1}=0.2$, $\lambda_{3}=0.9$, as $\epsilon$ is decreased from $1.0$ to $0.0$. (a) Stationary pulse patterns observed in region I: $\eps\in (0,0.32)$; (b) "Inside-group" zigzag patterns observed in region II: $\epsilon \in (0.33,0.64)$; (c) Rotating wave (traveling pulse) patterns observed in region III: $\epsilon \in (0.65,1)$.}
\label{Fig_Patterns_EpsilonTo0}
\end{figure}

Figure \ref{Fig_Amplitudes_Epsilon}(a) shows the amplitude of the patterns obtained when $\epsilon \in [0,1]$, for the particular parameter values mentioned before. Since some of these amplitudes show time-oscillations between different values, we graph the maximum and minimum values of these amplitudes for each $\epsilon$.
As we decrease $\epsilon$ from 1.0 towards 0.64 (region III), the amplitude undergoes some very small temporal oscillations (see also panel (c)). This amplitude corresponds to the rotating wave patterns (with a small time-modulation) shown in Figure \ref{Fig_Patterns_EpsilonTo0}(c). For $\epsilon \in (0.32,0.64)$ (region II), the amplitude oscillates between two large values. This corresponds to the ``inside-group" zigzagging behaviour shown in Figure \ref{Fig_Patterns_EpsilonTo0}(b) near $x=6$, where the group as a whole does not move in space but individuals inside the group move between the left and right edges of the group. We also note a period-doubling bifurcation at $\epsilon=0.61$, which leads to a slight decrease in the amplitude. Finally, as $\epsilon$ is decreased below 0.2 (region I), the movement inside the group is lost and the pattern is described by stationary pulses with fixed amplitude (see Figure \ref{Fig_Amplitudes_Epsilon}(a) and Figure \ref{Fig_Patterns_EpsilonTo0}(a)). 
Figures \ref{Fig_Amplitudes_Epsilon}(b),(c) show the time-variation of the amplitudes of the spatial and spatiotemporal patterns obtained for $\epsilon \in [0,1]$. Figures \ref{Fig_Amplitudes_Epsilon}(d)-(f) show the density profiles of the patterns observed in regions I-III. 

\begin{figure}[!ht]
\centering
 \includegraphics[width=5.4in]{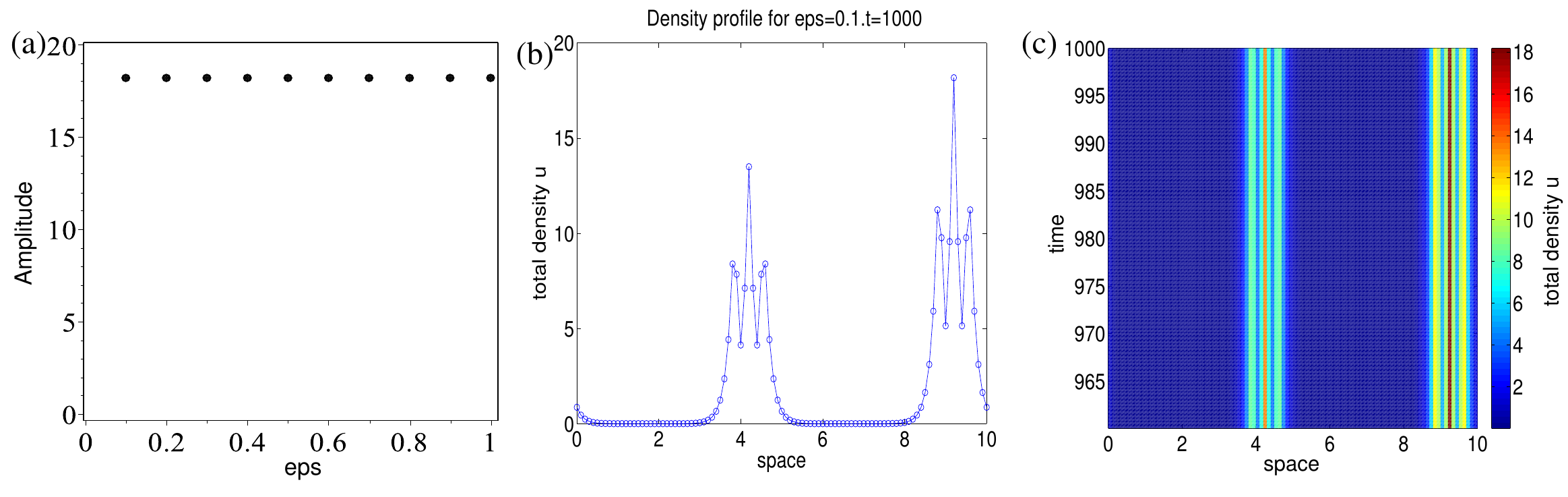}
\caption{The amplitude and density of the patterns obtained for model M2 with $q_{a}=2.2$, $q_{r}=0.93$, $q_{al}=0$, $\lambda_{1}=0.2$, $\lambda_{2}=0$, $\lambda_{3}=0.9$, as $\epsilon$ is decreased from $1.0$ to $0.0$.  (a) Bifurcation diagram for the amplitude of the patterns as a function of $\epsilon$. (b) Density profile for the stationary patterns. (c) Time-space plot of the density.}
\label{Fig_Amplitudes_Epsilon_M2}
\end{figure}

Because the macro-scale models ($\epsilon=0$) seem to exhibit stationary pulses (as shown in Figure \ref{Fig_Patterns_EpsilonTo0}(a)), we now start with these stationary pulses (for $\epsilon=1$) and investigate whether they change in any way as $\epsilon \to 0$. To this end, we focus on model  M2 (see Figure \ref{Models_M2_M4_Description}).
Figure \ref{Fig_Amplitudes_Epsilon_M2} shows the amplitude of the stationary pulses obtained with model M2 in a particular parameter region ($q_{a}=2.2$, $q_{r}=0.93$, $q_{al}=0$; see also Figure \ref{Fig_Ss_M4_Epsilon}), as we decrease the scaling parameter $\epsilon$. We observe that in this case, the scaling does not affect the patterns or their amplitudes. 

\section{Summary and Discussion}\label{Sect:discussion}

In this study, we investigated the connections between various 1D and 2D non-local kinetic and macroscopic models for self-organised biological aggregations. The non-locality of these models was the result of the assumptions that individuals can interact with neighbours positioned further away, but still within their perception range. To simplify the kinetic models that incorporate microscopic-level interactions (such as individuals' speed and turning rates), one can consider different scaling approaches, which transform these models into corresponding parabolic and hyperbolic models (described in terms of average speed and average turning behaviour). Here, we focused on three types of scalings namely, parabolic, hyperbolic and grazing collision limits. We showed that while for the kinetic models the non-local interactions influence the turning rates (i.e., individuals turn to approach their neighbours, to move away from them or to align with them), for the limiting parabolic and hyperbolic models the non-
local interactions influence the dispersion and the drift of the aggregations. In particular, we showed that the assumption that individuals can turn randomly following the non-directional perception 
of neighbours around them leads, in the macroscopic scaling, to density-dependent diffusion. Moreover, this diffusion decreased with the increase in the population density. Biologically, this means that larger animal groups are less likely to spread out.

Next, we investigated how two types of patterns (i.e., travelling and stationary aggregations) displayed by the 1D kinetic models, were preserved in the limit to macroscopic parabolic models. To this end, we first investigated the local stability of spatially homogeneous patterns characterised by individuals spread evenly over the domain, and showed that local Hopf bifurcations are lost in the parabolic limit. These Hopf bifurcations give rise to travelling aggregations (i.e., rotating waves). We then tested this observation numerically, with the help of asymptotic preserving methods. We started with a rotating wave pattern obtained near a Hopf/Steady-state bifurcation for $\epsilon=1$ (1D kinetic model), and studied numerically how does this pattern change when $\epsilon \to 0$ (1D parabolic model). By graphing the amplitude of the resulting patterns as the scaling parameter $\epsilon$ is decreased from $\epsilon=1$  to $\epsilon=0$, we showed that there were two major transitions. The first transition 
occurred around $\epsilon=0.64$, when the travelling (rotating) groups stopped moving. We note, however, that while the group as a whole was stationary, the individuals inside the group were still moving between the left- and right-edges of the group, leading to an ``inside-group" zigzagging behaviour. The second transition occurred around $\epsilon=0.32$, when the individuals inside the groups stopped moving, leading to stationary pulses. 

We emphasise here that this study is one of the first in the literature to investigate numerically the transitions between different aggregation patterns, as a scaling parameter $\epsilon$ is varied from values corresponding to mesoscale dynamics ($\epsilon=1$) to values corresponding to macroscale dynamics ($\epsilon=0$). Understanding these transitions is important when investigating biological phenomena that occurs on multiple scales, since it allows us to make decisions regarding the models that are most suitable to reproduce the observed dynamics.

In this study we investigated only the preservation of patterns via the parabolic limit. Similar investigations could have been performed for the hydrodynamic limit or the grazing collision limit. Moreover, we investigated only the bifurcation of two types of patterns displayed by model (\ref{1D_model}), namely travelling and stationary aggregations. However, as shown previously \cite{Eftimie2}, model (\ref{1D_model}) could display many more types of complex spatio-temporal patterns. We stress that our aim here was not to investigate how all types of possible patterns are preserved by all these different scaling approaches. Rather, it was to show that by taking these asymptotic limits,  some patterns could be lost. Therefore, even if the macroscopic models are simpler to investigate, they might not exhibit the same patterns as the kinetic models. We also tried to emphasise the usefulness of asymptotic preserving numerical methods to understand the bifurcation of the solutions as one investigates the 
transition from mesoscopic-level to macroscopic-level aggregation dynamics.

\section*{Acknowledgements}
JAC acknowledges support from projects MTM2011-27739-C04-02, 2009-SGR-345 from Ag\`encia de Gesti\'o d'Ajuts Universitaris i de Recerca-Generali\-tat  de Catalunya, the Royal Society through a Wolfson Research Merit Award, and the Engineering and Physical Sciences Research Council (UK) grant number EP/K008404/1. RE acknowledges support from an Engineering and Physical Sciences Research Council (UK) First Grant number EP/K033689/1. 
%

\end{document}